\documentclass[a4paper,11pt,nohyper]{JHEP3}
\usepackage{amsmath,latexsym,amssymb,slashed}
\usepackage{graphicx}
\usepackage{psfrag}
\usepackage[latin1]{inputenc}
\usepackage{subfigure}
\usepackage{nicefrac}
\usepackage{bbm}
\usepackage[stable]{footmisc}

\usepackage{comment}

\newcommand\beal{\begin{align}}
\newcommand\nn{\nonumber}

\newcommand{\eq}[1]{\begin{equation}#1\end{equation}}
\newcommand{\spl}[1]{\begin{split}#1\end{split}}
\newcommand{\al}[1]{\begin{align}#1\end{align}}

\newcommand{\arXividhepth}[1]{\href{http://arxiv.org/abs/#1}arXiv:{\tt #1} [hep-th]}

\newcommand{\ads}{AdS_4}

\newcommand{\mcal}{\mathcal{M}}
\newcommand{\ccal}{\mathcal{C}}
\newcommand{\ncal}{\mathcal{N}}
\newcommand{\boxedeq}[1]{
\begin{equation}
\fbox{
\rule[0.7cm]{0pt}{0pt}
$#1$
\rule[-0.45cm]{0pt}{0pt}
}
\end{equation}
}

\def\slashchar#1{\setbox0=\hbox{$#1$}           
\dimen0=\wd0                                 
\setbox1=\hbox{/} \dimen1=\wd1               
\ifdim\dimen0>\dimen1                        
\rlap{\hbox to \dimen0{\hfil/\hfil}}      
#1                                        
\else                                        
\rlap{\hbox to \dimen1{\hfil$#1$\hfil}}   
/                                         
\fi}

\title{Classes of AdS$_4$ type IIA/IIB compactifications\\ with
SU(3)$\times$SU(3) structure}

\author{Dieter L\"{u}st${}^{\diamondsuit\clubsuit}$ and 
Dimitrios Tsimpis${}^{\clubsuit}$  \\

\begin{itemize}
 
\item  Max-Planck-Institut f\"ur Physik\\
F\"ohringer Ring 6, 80805 M\"unchen, Germany
 
\item  Arnold-Sommerfeld-Center for Theoretical Physics\\
Department f\"ur Physik, Ludwig-Maximilians-Universit\"at M\"unchen\\
Theresienstra\ss e 37, 80333 M\"unchen, Germany
 \end{itemize}

\bigskip
E-mail:
\email{dieter.luest@lmu.de} \& \email{luest@mppmu.mpg.de}, \email{dimitrios.tsimpis@lmu.de}}

\abstract{
We introduce an ansatz which allows us to solve the supersymmetry equations for 
warped $\mathcal{N}=1$ $AdS_4$ type II supergravity compactifications of general $SU(3)\times SU(3)$ structure. 
As a byproduct we obtain a set of necessary conditions which 
every supersymmetric $AdS_4$ vacuum should obey. 
The case of $AdS_4$ compactifications of IIB on manifolds of 
static $SU(2)$ structure is examined in detail. 
Several examples of solutions are presented. In the limit 
of four-dimensional Minkowski space, we present examples of 
supersymmetric IIB warped compactifications with partially localized NS5- and D5-branes. 
We also present `massive' non-supersymmetric $AdS_4\times\mathcal{M}_6$ solutions of IIA,
where $\mathcal{M}_6$ can be any six-dimensional  Einstein-K\"{a}hler manifold. 
}

\preprint{LMU-ASC 05/09\\MPP-2009-14}

\keywords{Anti-de Sitter vacua, G-structures}

\begin{document}
\setcounter{footnote}{0}
\renewcommand{\thefootnote}{\arabic{footnote}}
\setcounter{section}{0}
\section{Introduction}
\label{introduction}

$AdS_4$ vacua of type IIA string theory are examples of flux vacua in which all moduli can be stabilized at tree level, in a regime where the quantum corrections to the supergravity approximation are parametrically small. As such they appear phenomenologically promising and can serve as a starting point for the construction, upon uplifting,  of 
metastable de Sitter vacua and models of inflation. 
Another strong motivation for the study of $AdS_4$ vacua is related to the $AdS_4/CFT_3$ duality and the 
recent progress in our understanding of the world-volume theory of coincident M2 branes \cite{blg, abjm}. It has been observed, 
however, that at the moment there are many more  three-dimensional superconformal field theories 
than there are examples of $AdS_4$ supergravity vacua in M-theory or IIA supergravity. 

All known examples to date of supersymmetric $AdS_4$ vacua of (massive) IIA fall in the general class of 
rigid $SU(3)$ solutions (an explanation of the terminology will follow shortly) given in \cite{lt}. This 
class includes the celebrated Nilsson-Pope $\mathcal{N}=6$ and $\mathcal{N}=1$ $AdS_4\times \mathbb{CP}^3$ vacua 
\cite{np,npa,npb} as limiting cases\footnote{The fact that the Nilsson-Pope solutions belong to the class of \cite{lt} was first pointed out in 
\cite{font}.}, the nearly-K\"{a}hler vacua 
of Behrndt-Cvetic \cite{bc}, as well as the vacua recently constructed by Tomasiello \cite{toma}. Finally, in \cite{klt}, 
all previously known  vacua, as well as some new ones, were constructed using left-invariant $SU(3)$ structures on groups and cosets. On the other hand, the type IIB side has been almost entirely  unexplored, perhaps due to a no-go theorem which forbids IIB $AdS_4$ vacua with $SU(3)$-structure \cite{bciib}. 
It is the purpose of this paper to go beyond the list of solutions in \cite{klt} and the analysis of \cite{bciib}, and 
take a step towards the construction of more general type II $AdS_4$ vacua. 

Supersymmetric solutions of type II supergravity of warped-product form: $AdS_4\times_{w} \mathcal{M}_6$, 
where $\mathcal{M}_6$ is the internal six-dimensional manifold, can be described in terms of two globally-defined  
internal spinors $\theta_{1,2}$ specifying the spinor ansatz of the solution. These two internal spinors must be of 
equal norm and proportional to the warp factor, as a consequence of supersymmetry\footnote{This 
was first observed in \cite{lt} in the special case of rigid 
$SU(3)$ structure. In the general case of $SU(3)\times SU(3)$ structure it was first shown in the appendix 
of \cite{scan}.}. Hence, provided
 the warp factor is nowhere-vanishing, both spinors must be nowhere-vanishing. Since with each of the two internal spinors we can associate an $SU(3)$ structure, we therefore have 
 a {\it global} $SU(3)\times SU(3)$ structure on $\mathcal{M}_6$. In particular it follows that 
there is a reduction of the structure group of  $\mathcal{M}_6$ to $SU(3)$ or a subgroup thereof\footnote{
Contrary to what is sometimes claimed in the literature, supersymmetry need  not in general imply the reduction of the structure group of the internal manifold. One example is  
 compactifications of eleven-dimensional supergravity to three-dimensional maximally-symmetric space \cite{tsim}.}.

The different types of solutions can be classified according to the relative angle of the two spinors. Here we follow 
the terminology of \cite{andriot,cetal}, according to which we distinguish the following subcases of 
$SU(3)\times SU(3)$ structure:
  \begin{itemize}
\item {\it strict $SU(3)$ structure}: $\theta_1$ and $\theta_2$ are parallel everywhere;
\item {\it static $SU(2)$ structure}: $\theta_1$ and $\theta_2$ are orthogonal everywhere;
\item {\it intermediate $SU(2)$ structure}: $\theta_1$ and $\theta_2$  are at a constant angle, which is neither zero nor a right angle;
\item {\it dynamic $SU(3)\times SU(3)$ structure}: the angle between $\theta_1$ and $\theta_2$  varies, possibly becoming zero or a right angle at special loci.
\end{itemize}

It was shown in \cite{blt} that there can be no IIA $AdS_4\times_{w} \mathcal{M}_6$ vacua of static $SU(2)$ 
structure\footnote{This no-go was subsequently generalized in \cite{cetal} to include {\it left-invariant} intermediate 
$SU(2)$ structure.}. As already mentioned, there is an analogous no-go theorem in IIB forbidding 
$AdS_4\times_{w} \mathcal{M}_6$ vacua of strict $SU(3)$ structure. 
To go beyond the static $SU(2)$ and strict $SU(3)$ structure cases, 
we must search for vacua of either dynamic $SU(3)\times SU(3)$ or intermediate $SU(2)$ structure.

\begin{figure}
\centering
\includegraphics[width=8cm]{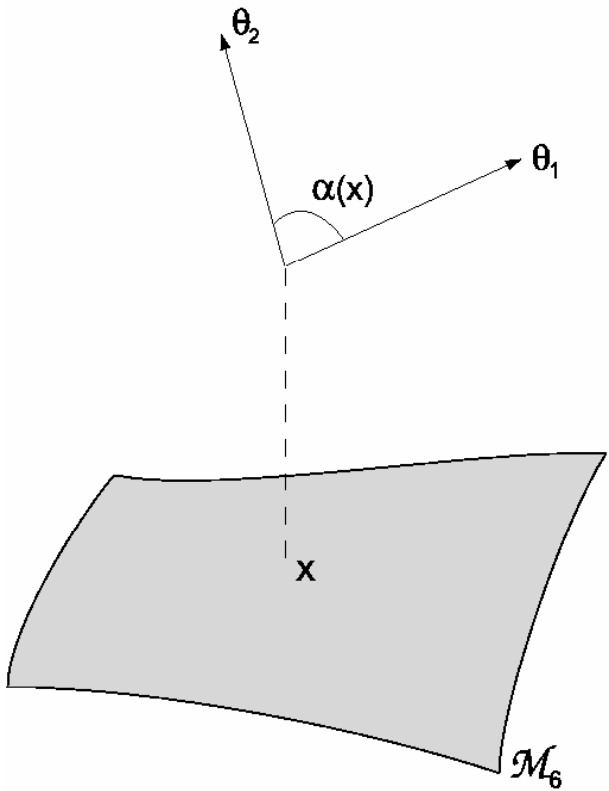}
\caption{The angle $\alpha(x)$ between the two internal spinors $\theta_{1,2}$ is, in general, 
 a function of the position $x\in\mathcal{M}_6$.}
\label{spinorsfigure}
\end{figure}

The supersymmetry equations of ten-dimensional 
type II supergravity for a generic global $SU(3)\times SU(3)$-structure ansatz can be elegantly
 formulated in the language of generalized geometry \cite{gene}. In searching for explicit examples 
of supersymmetric solutions, however, a different approach may be more promising: This is based on the observation 
that, assuming we do not have a rigid $SU(3)$ structure,  
 the two $SU(3)$ structures corresponding to each of the two internal spinors will generally interesect on a common 
$SU(2)$ subgroup. In other words, we can always define a preferred {\it local} $SU(2)$ structure on $\mathcal{M}_6$. 
Furthermore, we can expand all fluxes in terms of irreducible $SU(2)$ modules, upon which the 
analysis of the supersymmetry conditions reduces to a set of algebraic equations for the fluxes and the torsion 
classes of the local structure. 

The direct approach described in the preceding paragraph leads in general to cumbersome equations which cannot  
easily be solved, except of course in the case of rigid $SU(3)$ structure in IIA where several solutions are known by now. 
In order to make progress we need to look for further simplifications. In the present paper   
we propose the following rather natural ansatz: we demand that 
the representation-theoretic content of the solution consist entirely of scalars with respect to 
the local $SU(2)$ structure. In other words, in the decomposition of the various fluxes and torsion classes with 
respect to the local $SU(2)$ structure, we set to zero all components which are not scalar. In the following 
 we will refer to this as the {\it scalar ansatz}.

Imposing the scalar ansatz leads to considerable simplification, which enables us to explicitly solve the supersymmetry equations. The final result can be divided into two parts: (a) the part that constrains the fluxes, and (b) the part 
that specifies the local $SU(2)$ structure of the internal manifold. Part (a) of the solution is given below in 
eqs.~(\ref{susysol},\ref{constraint}) for IIA, and eqs.~(\ref{susysolb},\ref{constraintb}) for IIB.  
In both cases (\ref{4.0},\ref{3.8}) hold.

There is no obstruction to 
solving the equations specified in part (a): they simply express some of the flux components in terms of 
a set of free parameters. Moreover, 
these equations must be satisfied by all supersymmetric solutions, 
not only solutions obeying the scalar ansatz. In 
other words, they are necessary conditions for a supersymmetric $AdS_4$ vacuum; 
to our knowledge this is the first time they have been explicitly formulated.

Part (b) of the solution is 
given below in eq.~(\ref{23.1}), which is common to both IIA and IIB. 
Contrary to part (a) of the solution which is 
unobstructed, not every six-dimensional manifold will admit a local $SU(2)$ structure 
obeying (\ref{23.1}). Therefore, the reformulation of the supersymmetry equations in the language of the 
present paper provides a clear prescription for constructing new supersymmetric type II $AdS_4$ solutions: {\it scan 
for six-dimensional manifolds which admit a local $SU(2)$ structure obeying eq.~(\ref{23.1}). }

As is well-known, supersymmetry alone is not enough to guarantee that all equations of motion 
are satisfied, although it goes a long way. Even in the presence of calibrated (which in the present context 
can be taken to mean supersymmetric) sources, there is an integrability theorem 
which guarantees that, provided the Bianchi identities are satisfied\footnote{Here 
we adopt the terminology of the `democratic' formalism in which the (generalized) Bianchi identities 
of the RR fields also include the equations of motion in the traditional 
sense.}, all remaining equations 
of motion will be automatically satisfied \cite{kt}. In general the Bianchi identities will indeed include 
source contributions, which may or may not admit satisfactory physical interpretation. This analysis has to be performed 
in addition to the analysis of the supersymmetry equations.

The remainder of the paper is organized as follows: Section \ref{sec:supersymmetrya} introduces 
the scalar ansatz and presents the general solution to the supersymmetry equations. Section \ref{sec:iiaexamples} 
contains examples of IIA solutions. In particular, section \ref{sec:iiasmeared} contains examples 
of supersymmetric IIA solutions with smeared sources. Unfortunately these do no seem to admit a satisfactory 
physical interpretation.  
Section \ref{eomsolutions} contains a number of supergravity vacua of the form $AdS_4\times\mathcal{M}_6$, 
where $\mathcal{M}_6$ can be any six-dimensional Einstein-K\"{a}hler manifold. These solutions are shown
 to be non-supersymmetric, as they violate the necessary conditions of section \ref{sec:solution}. They are 
 anticipated already by Romans in \cite{roma}, although their existence is only mentioned very briefly in that reference (see 
the comment below eq.~(28) of \cite{roma}).

Section \ref{sec:staticsu2} analyzes in detail 
the special case of supersymmetric $AdS_4$ solutions of static $SU(2)$ structure. This case is, in a sense, 
the analogue of the strict $SU(3)$ case analyzed in \cite{lt}, however it had not been systematically 
analyzed before in the literature. The complete solution to the supersymmetry equations, subject to the 
scalar ansatz, is given in eqs.~(\ref{solstatic}-\ref{23.1iib}) below. Section \ref{sec:smearediib} 
contains two examples of solutions with smeared sources, which have appeared before in the literature, while 
section \ref{sec:iibloc} contains an example with partially-localized sources, which to our knowledge is new.

The appendices \ref{useful},\ref{tensordecompositions},\ref{sec:localstructure} contain useful relations and many technical details of the results presented
in the main text. Appendix \ref{sec:sasaki} reviews the relation between six-dimensional Einstein-K\"{a}hler and 
seven-dimensional Sasaki-Einstein  manifolds.

{\bf Note added:} Several months after section \ref{eomsolutions} of the present paper 
was completed, we received preprint \cite{gato} 
which has also independently arrived at the solutions presented in that section.

\section{Supersymmetry}\label{sec:supersymmetrya}

In this section we introduce in detail the {\it scalar ansatz} referred to in the introduction,  
and we present the solution, under this ansatz, to the supersymmetry equations 
for backgrounds of the form $AdS_4\times_w\mathcal{M}_6$. As a corollary we 
derive a set of necessary conditions (eqs.~(\ref{susysol},\ref{constraint}) for IIA 
and eqs.~(\ref{susysolb},\ref{constraintb}) for IIB) 
which must hold for any supersymmetric $AdS_4$ vacuum -- not only 
for vacua obeying the scalar ansatz. To our knowledge this is the first time these 
conditions explicitly appear in the literature.

We follow the conventions of \cite{lmmt}, which the reader may consult for further details.  
We perform a four-plus-six spacetime split, according to which the 
ten-dimensional metric takes the warped-product form:
\beal
ds^2=e^{2A(x)}ds^2_4+g_{mn}dx^mdx^n
~,
\end{align}
where $\exp\! A$ is the warp factor, $ds^2_4$ is the line element of $AdS_4$ and $g_{mn}$ is the internal-manifold metric. 
The type IIA supersymmetry parameter is decomposed accordingly as:
\beal
\epsilon_i=\zeta\otimes\theta_i+\mathrm{c.c.}~; ~~~~~i=1,2~,
\end{align}
where $\epsilon_{1,2}$ are positive-, negative-chirality ten-dimensional Majorana spinors, and 
$\theta_{1,2}$ are positive-, negative-chirality six-dimensional complex spinors.  
$\zeta$ is a four-dimensional positive-chirality Killing spinor obeying:
\beal
\nabla_{\mu}\zeta=\frac{1}{2}W^*\gamma_{\mu}\zeta^*~,
\end{align}
where $|W|$ is the inverse radius of curvature of $AdS_4$. Moreover we are using the democratic formalism in which the 
RR fluxes take the form: 
\al{\label{here}
F^{tot}=vol_4\wedge\tilde{F}+F
~,}
so that the self-duality condition reads $\tilde{F}=\star_{6}\sigma(F)$, where $\star_{6}$ is the 
Hodge-star on $\mcal_6$ and $\sigma$ reverses 
the order of the indices. 

With these ans\"{a}tze, 
the supersymmetry equations for type IIA/IIB can be cast in the form of a set of `algebraic' equations:
\beal\label{susyeqsalg}
0 &= \slashed\partial A	\theta_1-\frac14 e^{\phi}\slashed{F}\theta_2+e^{-A}W\theta_1^* \cr
0 &=  \slashed\partial A\theta_2-\frac14 e^{\phi}\gamma_7\slashed{F}^\dagger\theta_1+e^{-A}W\theta_2^* \cr
0 &= (\slashed\partial\phi-2\slashed\partial A+\frac12\slashed{H})\theta_1
+\frac18 e^\phi\gamma^m\slashed{F}\gamma_m\gamma_7\theta_2-2e^{-A}W\theta_1^* \cr
0 &=  (\slashed\partial\phi-2\slashed\partial A-\frac12\slashed{H})\theta_2
-\frac18 e^\phi\gamma^m\slashed{F}^\dagger\gamma_m\theta_1-2e^{-A}W\theta_2^* ~,
\end{align}
together with a pair of `differential' equations:
\beal\label{susyeqsdiff}
0 &= (\nabla_m+\frac14\slashed{H}_m)\theta_1+\frac18 e^\phi\slashed{F}\gamma_m\gamma_7\theta_2 \cr
0 &=  (\nabla_m-\frac14\slashed{H}_m)\theta_2-\frac18 e^\phi\slashed{F}^\dagger\gamma_m\theta_1
~,
\end{align}
where $\gamma_7$ is the chirality matrix in six dimensions. Moreover $\gamma_7\theta_1=\theta_1$ in both IIA/IIB,  
while $\gamma_7\theta_2=-\theta_2$ in IIA and $\gamma_7\theta_2=\theta_2$ in IIB.

\subsection*{Local $SU(2)$ structure}

For the analysis of the supersymmetry it will be useful to work with a local  basis of 
orthogonal unimodular spinors 
$\eta_{1,2}$, with respect to which we can parameterize: 
\boxedeq{\label{3.5}
\theta_1=a~\eta_{1};~~~~~\theta_{2}=\left\{ 
\begin{array}{lll} b~\eta^*_{2}+c^*\eta^*_{1} & ~~~&\mathrm{IIA} \\
 b~\eta_{2}+c~\eta_{1}  &~~~ & \mathrm{IIB}
\end{array}\right.
~.
}
We can take $a,b\in\mathbb{R}$, by making 
 use of the freedom in the definition of the phase of $\eta_{1,2}$, while generally  $c\in\mathbb{C}$. 
This is the most general spinor ansatz, and is related to the `dielectric spinors' of \cite{zafa, hato, andriot}. In the context of $AdS_4$ compactifications of IIA, 
the two limiting cases $b=0$, corresponding to rigid $SU(3)$ structure, and $c=0$, 
corresponding to static $SU(2)$ structure, were considered in \cite{lt,blt} respectively. 
The most general spinor ansatz (\ref{3.5}) has not been analyzed before 
in this context\footnote{See \cite{andrun} for certain 
dynamic $SU(3)\times SU(3)$ IIA/IIB ans\"{a}tze, which however
do not seem to lead to solutions.}, although it is of course implicit in the 
generalized-geometry formulation of \cite{gene}.

The spinors $\theta_{1,2}$ define a (dynamic, in general) $SU(3)\times SU(3)$ structure, whereas the spinors $\eta_{1,2}$ define locally a static $SU(2)$ structure.  
The particular parametrization of $\theta_{1,2}$ in terms of $\eta_{1,2}$
above is chosen to be valid {\it a priori} on open patches where $\theta_1$ is non-vanishing\footnote{This can be seen by `inverting' (\ref{3.5}) 
to get
\beal
a=|\theta_1|~;~~~~~ b=\frac{1}{|\theta_1|}\sqrt{|\theta_1|^2|\theta_2|^2-|\theta_1\cdot \theta_2|^2}~;~~~~~c^*
=\frac{\theta_1\cdot \theta_2}{|\theta_1|} 
~.\nn
\end{align}
}.
However, as already mentioned in the introduction, $\theta_{1,2}$ are nowhere-vanishing 
hence this requirement is automatically satisfied (see the discussion immediately below
 eq.~(\ref{3.8})).

Each of the two orthogonal spinors defines an $SU(3)$ structure: 
\al{\label{3.54}
J^{(r)}_{mn}&:=i\eta_{r}\gamma_{mn}\eta^*_{r}\nn\\
\Omega^{(r)}_{mnp}&:=\eta_{r}\gamma_{mnp}\eta_{r}~, 
}
for $r=1,2$. 
The local static $SU(2)$ structure $(\widetilde{J},\omega)$ is the `intersection' of these two $SU(3)$ structures. It can be expressed 
in terms of $(J^{(r)}, \Omega^{(r)}, K)$, where $K$ is a holomorphic one-form given by 
\al{\label{3.55}
K_m:=\eta_{2}\gamma_m\eta_{1}~. 
}
As can be seen from (\ref{3.5}), the additional information contained in the one-form $(b/a)K_m$, 
which {\it e.g.} in IIA  is proportional to $(\theta_2^*\gamma_m\theta_1)/|\theta_1|^2$, can be thought of as 
parametrizing the deviation of the spinor ansatz from the rigid-$SU(3)$ case. Specifically:
\begin{align}\label{2.87}
J^{(1)}=\frac{i}{2}K\wedge K^*+ \widetilde{J}~&; ~~~~~J^{(2)}=\frac{i}{2}K\wedge K^*- \widetilde{J}\nn\\
\Omega^{(1)}=-i{\omega}\wedge K~&;~~~~~\Omega^{(2)}=i{\omega}^*\wedge K~,
\end{align}
where $\iota_K\widetilde{J}$, $\iota_K\omega$, $\iota_{K^*}\omega=0$. 
Moreover we have:
\beal\label{3.6}
{\omega}_{mn}:=i\eta_1\gamma_{mn}\eta_2^*~.
\end{align}

To analyze the content of supersymmetry, we will make repeated use of 
a number of additional identities satisfied by $\eta_{1,2}$ and the various 
forms introduced above. These can be found in \cite{blt}, whose  
spinor notations and conventions we follow\footnote{Unlike in \cite{blt}, in the present paper 
we do  not use superspace conventions for the forms.}.

\subsubsection*{Scalar ansatz}

The scalar ansatz proposed in the present paper consists of the following rather natural
 simplification:  we demand that in the decomposition of the various fluxes with 
respect to the local $SU(2)$ structure all components which are not scalar be set to zero.

Imposing the scalar ansatz, {\it i.e.} keeping only the scalars in the tensor decompositions 
given in appendix \ref{tensordecompositions}, leads 
to considerable simplification upon which the various RR forms read, in form-notation:
\beal\label{b16}
e^{\phi}F_0&=f_0\nn\\
e^{\phi}F_2&=\frac18 \left(
 f_2~\!\omega^*+ f_3\widetilde{J} +2i f_1K\wedge K^*\right)
+\mathrm{c.c.}\nn\\
e^{\phi}F_4&=
\frac{1}{16}g_1\widetilde{J}\wedge \widetilde{J}
+\frac{i}{96}\left(g_2~\!\omega^*+g_2^*~\!\omega +2g_3\widetilde{J}  
\right)\wedge K\wedge K^*\nn\\
e^{\phi}F_6&=f~\! vol_6
~,
\end{align}
for type IIA, while:
\beal\label{b17}
e^{\phi}F_1&=g_1K+\mathrm{c.c.}\nn\\
e^{\phi}F_3&=\frac{1}{24}\left(
f_1\omega^* +f_2~\!\omega+2f_3\widetilde{J} 
\right)\wedge K
+\mathrm{c.c.}\nn\\
e^{\phi}F_5&=g_2\star_6\! K+\mathrm{c.c.}
~,
\end{align}
for type IIB. 
In IIA the scalars $f$, $f_{0,1,3}$, $g_{1,3}$  are real, while $f_2$, $g_2$ are complex. 
In IIB all five scalars $f_{1,2,3}$, $g_{1,2}$ are complex.  
Note that in both cases the decompositions are parameterized by 
five complex scalar degrees of freedom. 
The expansion for the NSNS three-form is the same in both IIA, IIB:
\beal\label{b18}
H&= \frac{1}{24}\left(
h_1\omega^* +h_2~\!\omega+2h_3\widetilde{J} 
\right)\wedge K
+\mathrm{c.c.}
~,
\end{align}
where the scalars $h_{1,2,3}$ are complex.


\subsection{IIA solution}\label{sec:solution}

Plugging the expressions for the form fields (\ref{b16},\ref{b18}) into the algebraic suspersymmetry equations (\ref{susyeqsalg}) above and projecting onto the singlet of the local $SU(2)$ structure, we 
%
%
obtain the 
following solution:
%
%
%
\boxedeq{\label{susysol}
\spl{
f&=-3~\!\mathrm{Im}\left(\frac{c}{a}W\right)e^{-A} \\
\frac{{c}}{{b}}g_2&=g_3-6if_3-48i~\!\mathrm{Im}\left(\frac{c}{a}We^{-A}-\frac{b}{a}K\cdot\partial A\right) \\
g_1&=8f_0-\frac{2}{3}g_3-32~\!\mathrm{Re}\left(\frac{c}{a}We^{-A}+\frac{b}{a}K\cdot\partial A\right)\\
f_1&=-\frac12f_3-\mathrm{Im}\left(\frac{c}{a}We^{-A}-\frac{4b}{a}K\cdot\partial A\right) \\
\frac{{c}}{{b}}f_2&=f_3+\frac{i}{6}g_3+8i~\!\mathrm{Re}\left(\frac{c}{a}We^{-A}+\frac{b}{a}K\cdot\partial A\right)
-\frac{8ia}{b}K^*\cdot\partial A\\
\frac{a}{{b}}h_3&=\frac32f_3-6\mathrm{Im}\left(\frac{c}{a}W\right)e^{-A}-12i\mathrm{Re}\left(\frac{c}{a}W\right)e^{-A}\\
&~~~~~~~~~~~~
+6if_0-\frac{i}{4}g_3+\frac{6ia}{b}K^*\cdot\partial( 3A-\phi)
-\frac{12ib}{a}K^*\cdot\partial A\\
\frac{{c}^*}{a}h_2&=-\frac{i}{4}g_3-18i~\!\mathrm{Re}\left(\frac{c}{a}W\right)e^{-A}
-\frac{i|{c}|^2}{a{b}}\mathrm{Im}h_3+\frac{{b}}{a}\mathrm{Re}h_3\\
&~~~~~~~~~~~~
+\frac{6ia}{b}\mathrm{Re}K\cdot\partial( 3A-\phi)-\frac{6ib}{a}K^*\cdot\partial(3A-\phi)
\\
h_1&=h_2^*-\frac{2i{c}^*}{{b}}\mathrm{Im}h_3
-\frac{12c^*}{b}\mathrm{Im}K\cdot\partial (3A-\phi)
~,
}}
%
where we have chosen the inverse $AdS_4$ radius $W$, the dilaton and warp factor $\phi$, $A$,  
and the scalars $f_0,f_3,g_3$ (see eq.~(\ref{b16})) 
as independent variables. Moreover we have defined $K\cdot\partial:=K^m\partial_m$, 
so that $K\cdot\partial S=\mathcal{L}_KS$. (We use the same notation both for the one-form $K_mdx^m$ and 
the vector $K^m(\partial/\partial x^m)$ obtained by raising the covariant index with the unique metric compatible with the  
$SU(3)\times SU(3)$ structure). The Romans mass 
is in general nonzero and enters the above equations 
via $f_0:=e^{\phi}F_0$. 

The equations above must hold for any supersymmetric IIA $AdS_4$ vacuum -- not only 
for vacua obeying the scalar ansatz. To our knowledge, this is the first time 
they appear explicitly in the literature. 
In addition to these equations one would in general have a number of non-scalar equations, 
{\it i.e.} those which are obtained by projecting the supersymmetry equations 
onto irreducible representations which are not singlets under the local $SU(2)$ structure. 
In the present case, these 
will turn out to be equivalent to (\ref{constraint},\ref{23.1}) below,  as a consequence of the scalar ansatz.


In addition to the equations above, the fact that $\eta_{1,2}$ are unimodular imposes the constraints:  
$\partial(\eta_{i}^{\dagger}\eta_{i})=0$, for $i=1,2$. There is one more constraint, 
$\partial(\eta_{1}^{\dagger}\eta_{2})=0$, which is a consequence 
of the orthogonality of $\eta_{1,2}$. Using the differential equations (\ref{susyeqsdiff}), it can be 
%
%
seen that these three constraints are equivalent to the following:
\boxedeq{\label{constraint}
\spl{
b~\! We^{-A}&=\frac{c^*}{2}~\!\mathrm{Re}K\cdot\partial\left( \log\frac{c^*}{a}+3A-\phi    \right) \\
0&=\mathrm{Im}K\cdot\partial\left( \log\frac{c^*}{a}+3A-\phi    \right)\\
a&=\mathrm{constant}\times e^{~\frac{1}{2}A}
~,
}}
%
together with:
\boxedeq{\label{4.0}
dS=\frac{1}{2}K^* \left(K\cdot\partial S\right)+\frac{1}{2}K \left(K^*\cdot\partial S\right)~,
}
where $S(x)$ is any one of the scalars $A$, $\phi$, $a$, $b$, $c$, and $x$ is the coordinate of $\mathcal{M}_6$.

Before we proceed, let us make a couple of comments 
about eqs.~(\ref{constraint},\ref{4.0}). 
It follows from the first two lines of (\ref{constraint}) that:
\beal
\mathrm{Re}K~\! W&=\frac{a}{2b}e^{-2A+\phi}d\left( e^{2A-\phi}\theta_1\cdot \theta_2    \right)
~,
\end{align}
where we have taken footnote 7 into account together with (\ref{4.0}) and the last line of (\ref{constraint}). The 
no-go theorem of \cite{blt} then follows immediately from the above, since $\theta_1\perp\theta_2$ 
implies $K\neq0$ and  $W=0$. 
The more general no-go of \cite{cetal} also follows similarly. 
Moreover, as was remarked in that reference, the way to circumvent the 
no-go would be to allow for $e^{2A-\phi}\theta_1\cdot \theta_2$ to vary over the internal manifold.

To  gain insight into the meaning of equation (\ref{4.0}), note that, as explained in more detail in \cite{blt}, 
$K$ can be used to define an almost product structure on $\mathcal{M}_6$. Consequently, the internal metric can 
locally be cast in the form:
\beal\label{4.1.1}
ds_6^2=\sum_{i,j=1}^4\tilde{g}_{ij}(x)dx^i\otimes dx^j+K\otimes K^*
~,
\end{align}
where
\beal
\mathrm{Re}K=\Phi(x)\left(dx^5+\sum_{i=1}^4\mathcal{A}_i(x)dx^i\right)~;
~~~~~\mathrm{Im}K=\Psi(x)\left(dx^6+\sum_{i=1}^4\mathcal{B}_i(x)dx^i\right)~.
\end{align}
Since $\tilde{g}_{ij}$, $\Phi$, $\Psi$, 
$\mathcal{A}_i$, $\mathcal{B}_i$ depend in general on all coordinates of $\mathcal{M}_6$, it follows  that 
(\ref{4.1.1}) is not in general a fibration. Condition (\ref{4.0}) can then locally be rewritten as:
\beal
e_a{}^m\frac{\partial}{\partial x^m}S=0~;~~~~~a=1,\dots , 4~.
\end{align}

Finally, 
in order to allow for AdS$_4$ solutions, $W\neq0$, 
it turns out that $a$, ${b}$, ${c}$ must satisfy the following relation: 
\boxedeq{\label{3.8}
a^2={b}^2+|c|^2~.
}
%
Equivalently, the measures of the two spinors $\theta_{1,2}$ must be equal:
\beal\label{3.8.1}
|\theta_1|^2=|\theta_2|^2~.
\end{align}
As already mentioned in the introduction, 
it follows from (\ref{3.8}), or equivalently (\ref{3.8.1}),
 and the last equation in (\ref{constraint}) that $\theta_{1,2}$ must be nowhere-vanishing. 
We therefore have a {\it globally} well-defined $SU(3)\times SU(3)$ structure on $\mathcal{M}_6$.

%

%
%

It is straightforward to verify that the results of \cite{blt} are  
recovered in the ${c}\rightarrow0$ limit, which corresponds to the static $SU(2)$ case. The limit $b\rightarrow0$, 
which corresponds to the strict $SU(3)$ case \cite{lt},  
can also be taken but is slightly more subtle, as in this limit the irreducible representations which appear 
in the tensor decompositions of the various fields, have to be taken with respect to the 
$SU(3)$ structure.

\subsection{IIB solution}

Proceeding similarly to  the IIA  case, taking eqs.~(\ref{b17},\ref{b18}) into account,  
the algebraic supersymmetry equations  (\ref{susyeqsalg}) can be solved to give:
%
%
%
\boxedeq{\label{susysolb}
\spl{
f_1&=12i\left\{ \frac{c}{b} (g_1+i g_2)+\frac{c}{a}W^*e^{-A}+(\frac{2a}{b}-\frac{b}{a})K^*\cdot\partial A \right\} \\
f_2&=12i\left\{ -\frac{c}{a}W^*e^{-A}+\frac{b}{a}K^*\cdot\partial A \right\}\\
f_3&=12i\left\{ \frac12 (g_1+i g_2)+\frac{b}{a}W^*e^{-A}-\frac{c}{a}K^*\cdot\partial A \right\}\\
h_1&=12i\left\{ (\frac{a}{b}-\frac{b}{2a}) (g_1-i g_2)-W^*e^{-A}
+\frac{c}{b} K^*\cdot\partial\left( 2A-\phi \right)\right\} \\
h_2&=12i\left\{ \frac{b}{2a} (g_1-i g_2)-W^*e^{-A}\right\}\\
h_3&=6i\left\{ -\frac{c}{a} (g_1-i g_2)
+K^*\cdot\partial\left( 2A-\phi \right)\right\}\\
\mathrm{Re}~\!c&=0
~.
}}
%
Note that the solution leaves the complex scalars $g_1$, $g_2$ unconstrained. 
Moreover, the constraints $\partial(\eta_i^{\dagger}\eta_j)=0$ imply:
\boxedeq{\label{constraintb}
\spl{
b~\! We^{-A}&=\frac{c}{3}K\cdot\partial\left( \phi-4A-\log\frac{|c|}{a}  \right) + \frac{2ia}{3}g_2^*\\
a&=\mathrm{constant}\times e^{~\frac{1}{2}A}
~.
}}
%
The equations above must hold for any supersymmetric IIB $AdS_4$ vacuum -- not only 
for vacua obeying the scalar ansatz. To our knowledge, this is the first time 
they appear explicitly in the literature. In addition, eqs.~(\ref{4.0},\ref{3.8}) hold in the present case as well.

\subsection{Local SU(2) structure}

The local $SU(2)$ structure of the internal manifold, encoded in the 
action of the exterior differential on $(K,\widetilde{J},\omega)$, can be read off  
using the differential supersymmetry equations (\ref{susyeqsdiff}) as explained in 
appendix \ref{sec:localstructure}. More specifically, for both IIA and IIB we 
can give the following compact 
expressions:
\boxedeq{\label{23.1}
\spl{
dK&=K^*\wedge K\left\{\frac{1}{2}(K^*)_1-\frac{1}{2}(K)^*_1-(K^*K)_2 +(KK)^*_2\right\}\\
&~~~~~~~~~~~~~~~
+\omega\left\{-4(\omega)_2\right\}+\omega^*\left\{2(\widetilde{J})_1\right\}+\widetilde{J}\left\{-2(\widetilde{J})_2-4(\omega)_1\right\}\\
d\tilde{J}&=K\wedge\omega\left\{-2(K^*K)^*_1-i(\widetilde{J})^*_1 \right\}+K\wedge\omega^*\left\{-2(KK)_1-2i(\omega)^*_2 \right\}\\
&~~~~~~~~~~~~~~~+K\wedge\widetilde{J}\left\{2i(\omega)^*_1-i(\widetilde{J})^*_2 \right\}+\mathrm{c.c}\\
d\omega&=
K\wedge\widetilde{J}\left\{4(KK)_1+4i(\omega)_2^*\right\}
+K^*\wedge\widetilde{J}\left\{4(K^*K)_1-2i(\widetilde{J})_1\right\}\\
&+K\wedge\omega\left\{\frac12(K)_1-\frac12(K^*)^*_1-(K^*K)^*_2+(KK)_2-2i(\widetilde{J})^*_2\right\}\\
&+K^*\wedge\omega\left\{\frac12(K^*)_1-\frac12(K)^*_1+(K^*K)_2-(KK)^*_2  -4i(\omega)_1 \right\}
~,
}}
which can be derived from (\ref{23},\ref{24.1},\ref{25.1}) with the use of (\ref{pirto}). All coefficients  
on the right-hand sides above are known and are explicitly given in eqs.~(\ref{21}-\ref{22b}). 
Since the geometry is determined by the local $SU(2)$ structure, eqn.~(\ref{23.1}) fixes the 
geometry in terms of the flux parameters.

Note that the 
local $SU(2)$ structure can also be specified either by the triplet $(K,{J}^{(1)},\Omega^{(1)})$, 
or, equivalently, $(K,{J}^{(2)},\Omega^{(2)})$. In the former case
 (\ref{23.1}) would have to be replaced by the epression for $dK$ (the first of the 
equations above) together with the expression for the torsion classes, given in (\ref{torcla}), for the $SU(3)$ structure 
corresponding to $({J}^{(1)},\Omega^{(1)})$. As already remarked below (\ref{3.55}), the additional information 
contained in the one-form $(b/a)K$ can be thought of as 
parametrizing the deviation of the spinor ansatz from the rigid-$SU(3)$ case.

{\bf In summary:} {\it for a supersymmetric background  
of the form $AdS_4\times \mathcal{M}_6$, the internal 
manifold $\mathcal{M}_6$ is specified by a 
local $SU(2)$ structure $(K,\widetilde{J},\omega)$ obeying (\ref{23.1}); 
the fluxes are given by (\ref{susysol},\ref{constraint}) in IIA, and 
by (\ref{susysolb},\ref{constraintb}) in IIB; in both cases (\ref{4.0},\ref{3.8}) hold. }

\section{IIA Examples}\label{sec:iiaexamples}

The reformulation of the supersymmetry equations in the present language readily suggests a strategy for a systematic search for solutions: Given an $SU(3)$-structure manifold $\mathcal{M}_6$ choose a family of triplets 
$(K^{\lambda},\widetilde{J}^{\lambda},\omega^{\lambda})$ on it, where $\lambda$ parameterizes the family;  
impose eqs.~(\ref{23.1}) in order to restrict $\lambda$; if a solution exists on $\mathcal{M}_6$, 
read off the fluxes using (\ref{susysol},\ref{constraint}). The following examples will illustrate this method for type IIA. In the next section we will consider the case of static $SU(2)$ structure in IIB.

\subsection{Examples with smeared sources}\label{sec:iiasmeared}

The following is a simple solution of the supersymmetry equations. 
Let us demand that $d\omega$ should not contain any $K\wedge\widetilde{J}$, $K^*\wedge\widetilde{J}$ terms. This can be seen from (\ref{23.1}) to automatically imply that $d\widetilde{J}$ contains only $K\wedge\widetilde{J}$, $K^*\wedge\widetilde{J}$ terms. In addition, we demand that $dK$ be proportional to $K^*\wedge K$, 
and that $K\cdot\partial A=0$. 
Taking the constraints (\ref{constraint}) into account, the aforementioned conditions imply:
\beal\label{simple1}
f_0&=\frac{4b^2+5|c|^2}{2ab}C~;~~~~f_1=0~;~~~~~f_2=\frac{2ic^*}{a}C~;~~~~~f_3=0\nn\\
g_1&=\frac{12a^2+4b^2}{ab}C~;~~~~~g_2=\frac{36c^*}{a}C~;~~~~~g_3=\frac{36|c|^2}{ab}C~\nn\\
h_1&=0~;~~~~~h_2=-\frac{12ic}{b}C;~~~~~h_3=6iC\nn\\
\mathrm{Im}&(cW)=0~; ~~~~~\mathrm{Im}K\cdot\partial\phi=0~; ~~~~~\mathrm{Re}K\cdot\partial\phi=C~; 
~~~~~a,~b,~c,~A=\mathrm{constant}~,
\end{align}
where we have introduced the real constant $C:=-(2b/c^*)e^{-A}W$. It readily  follows 
from the above that we have an intermediate $SU(2)$ structure.

In form notation the fluxes read:
\beal\label{30}
H&=\frac{i}{2}C\left(\widetilde{J}-\frac{c}{b}\omega^*\right)\wedge K+\mathrm{c.c.}\nn\\
e^{\phi}F_0&=\frac{4b^2+5|c|^2}{2ab}C~;~~~~~
e^{\phi}F_2=-\frac{ic}{4a}C^*\omega+\mathrm{c.c.}\nn\\
e^{\phi}F_4&=\frac{3a^2+b^2}{4ab}C\widetilde{J}\wedge\widetilde{J}
+\frac{3i}{4a}\left( \frac{|c|^2}{b}C\widetilde{J}+\mathrm{Re}(cC^*\omega) \right)K\wedge K^*~.
\end{align}
{}Furthermore, we can compute the local structure from (\ref{23.1}):
\beal\label{26.1}
d&\mathrm{Re}K=0~;~~~~~ d\big(e^{\phi}\mathrm{Im}K\big)=0\nn\\
d&\big(e^{-\phi}\omega\big)=0~;~~~~~ d\big(e^{-\phi}\widetilde{J}\big)=0~.
\end{align}
The above relations imply that $K$ can be written as $K=d\varphi+ie^{-\phi}d\chi$ for some 
local coordinates $\varphi$, $\chi$. It then follows from (\ref{simple1}) that the dilaton is given by 
\beal\label{27.1}
\phi=C(\varphi-\varphi_0)~,
\end{align}
for some constant $\varphi_0$. Moreover, as can be seen from (\ref{26}), the two-forms $\big(e^{-\phi}\omega\big)$, 
$\big(e^{-\phi}\widetilde{J}\big)$ define a four-dimensional Calabi-Yau manifold, {\it i.e.} a $K3$ surface. 
The metric of the six-dimensional internal manifold can therefore be written as:
\beal\label{28}
ds^2_6=e^{\phi}ds^2_{{K3}}+d\varphi^2+e^{-2\phi}d\chi^2~,
\end{align}
where  $ds^2_{{K3}}$ is the mertic of the $K3$ surface. Note that $\varphi$, $\chi$ parameterize 
a two-dimensional hyperbolic space $H_2$. 

Although the supersymmetry equations can be solved in the way 
described above, it is not difficult to see that 
the sourceless Bianchi identities cannot be satisfied for all form fields. 
In particular, negative-tension (non-localized) sources must be added, which is physically unsatisfactory. 
Although we will not list the details here, similar solutions of the supersymmetry 
equations (but not of the sourceless Bianchi identities) 
can be achieved by taking the internal manifold to be a nilmanifold. It is possible that performing 
a systematic scan of the nilmanifolds, something which we have not done,
 would yield supersymmetric solutions which also satisfy the 
sourceless Bianchi identities.


\subsection*{Constant warp factor, dilaton}

In the case of constant dilaton and warp factor, 
a simple way to solve (\ref{constraint},\ref{3.8}) is by making the following ansatz:
\beal\label{4.1}
b=a \cos\varphi;~~~~~c=a e^{i\delta}\sin\varphi;~~~~~W=|W|e^{-i\delta};~~~~~
\phi,~A,~a,~\delta=\mathrm{constant}~,
\end{align}
where we have parameterized:
\beal\label{4.2}
\mathrm{Re}K=\frac{e^A}{2|W|}d\varphi+\mathcal{A}~,
\end{align}
for some co-ordinate $\varphi$ and a one-form $\mathcal{A}$ such that 
$\iota_{\partial/\partial\varphi}\mathcal{A}=0$. In order to see  
that (\ref{4.1}) is indeed a solution of (\ref{constraint},\ref{3.8}), note that (\ref{4.2}) implies 
$\mathrm{Re}K\cdot\partial=2|W|e^{-A}\partial/\partial\varphi$.

\subsection{Examples without sources}\label{eomsolutions}

We will now consider a certain class of IIA 
compactifications of the form $AdS_4\times \mathcal{M}_6$, where  $\mathcal{M}_6$ can be  
any Einstein-K\"{a}hler manifold. We will allow for non-zero Romans mass, 
therefore these compactifications do not, in general, admit an eleven-dimensional lift.
 These solutions were anticipated by Romans in \cite{roma} (see also \cite{neu}), although their existence was 
only mentioned very briefly in that reference ({\it cf.} 
the comment below eq.~(28) of \cite{roma}).


For non-vanishing Romans mass these solutions will be shown, at the 
end of the present section, to be { non-supersymmetric}, 
as they do not obey the necessary 
supersymmetry conditions of section \ref{sec:solution}. On the 
other hand, for vanishing Romans mass we have an { enhancement 
of supersymmetry}, and the solutions fall within the class of 
the supersymmetric solutions of \cite{lt}. 
Using the known results, summarized 
in section \ref{sec:sasaki},  relating   six-dimensional Einstein-K\"{a}hler manifolds 
 to seven-dimensional Sasaki-Einstein manifolds, for vanishing Romans mass  
these solutions lift to the well-known {supersymmetric} 
M-theory solutions of Freund-Rubin type of the form $AdS_4\times \mathcal{M}_7$,  
where $\mathcal{M}_7$ is Sasaki-Einstein.

We take the ten-dimensional metric to be of the form:
\al{\label{2.1}
ds^2=ds^2(AdS_4)+ds^2(\mcal_6)
~,}
{\it i.e.} a direct (not warped) product $AdS_4\times \mathcal{M}_6$. Moreover, we take the NSNS three-form to vanish, $H=0$, 
and the dilaton to be constant. The RR fields are given by:
\al{\label{2.2}
F_{0}=\alpha;~~F_{2}=\beta J;~~F_{4}=\frac{1}{2}\gamma J^2;~~F_{6}=\frac{1}{6}\delta J^3
~,}
where $J$ is the K\"{a}hler form on $\mcal_6$, and $\alpha,\dots, \delta\in \mathbb{R}$.  
After imposing the 
self-duality condition, see below eq.~(\ref{here}),  the RR fluxes can be written more conventionally as:
\al{\label{2.3}
F^{tot}_{0}=\alpha;~~F^{tot}_{2}=\beta J;~~F^{tot}_{4}=\frac{1}{2}\gamma J^2+\delta~ \! vol_4~.
}
The following calculations are very similar to section 11.4 of \cite{lmmt}, so here we will simply state 
the results.

The NSNS Bianchi identity, $dH=0$, is trivially satisfied for this ansatz. Similarly, the 
generalized Bianchi identities for the RR fields, $d_HF=0$, (which in the conventional type II supergravity formulation 
correspond to both the Bianchi identities and the equations of motion) are also automatically satisfied 
by virtue of the closure of the K\"{a}hler form, $dJ=0$. It remains to examine the NS-sector equations of motion. 
The $H$-field equation of motion reduces to 
\al{\label{2.5}
\alpha\beta+2\beta\gamma+\gamma\delta=0
~.}
The dilaton equation reads:
\al{\label{2.6}
|W|^2-\frac{5}{8}\omega^2=0
~,}
where $W\in\mathbb{C}$, $\omega\in\mathbb{R}$ are related to the  
curvature of $AdS_4$, $\mcal_6$ via
\al{\label{2.7}
R_{\mu\nu}=-3g_{\mu\nu}|W|^2, ~~~~~R_{mn}=\frac{5}{4}\omega^2g_{mn}
~,}
respectively. 
Finally,  the external and internal Einstein equations read:
\al{\label{2.8}
|W|^2-\frac{1}{12}(\alpha^2+3\beta^2+3\gamma^2+\delta^2)=0
}
and
\al{\label{2.9}
5\omega^2+\alpha^2+\beta^2-\gamma^2-\delta^2=0
~,}
respectively. 

The full set of supergravity equations of motion above can be seen to admit three infinite classes of solutions. In 
each of these three classes, the constants $|W|$, $\omega$ can be solved for in terms of the 
real parameters $\alpha,\dots, \delta$ using (\ref{2.6},\ref{2.8}). 
Moreover we have:

{\it First solution:}
\boxedeq{\label{2.10}
\beta=\gamma=0~; ~~~~~\delta=\pm\sqrt{5}\alpha~.
}

{\it Second solution:}
\boxedeq{\label{2.11}
\alpha=\pm\frac{7}{5\sqrt{5}}\beta~;~~~~~ \gamma=\pm\frac{1}{\sqrt{5}}\beta~;~~~~~
\delta=-\frac{17}{5}\beta
~.}

{\it Third solution:} $\beta^2\geq3\gamma^2$ and
\boxedeq{\label{2.12}
\alpha=\gamma~\!\frac{-2\beta^2\pm\sqrt{(\beta^2-3\gamma^2)(9\beta^2+5\gamma^2)}}{\beta^2-5\gamma^2}~;~~~~~
\delta^2={5\alpha^2+9\beta^2+3\gamma^2}
~.}

The Nilsson-Pope `Hopf-fibration' solution \cite{np} is a subset of the third solution above, and is obtained upon setting the 
Romans mass to zero, $\alpha=0$. In this case we obtain:

{\it Hopf-fibration solution:}
\boxedeq{\label{2.13}
\alpha=\gamma=0~;~~~~~
\delta=\pm 3\beta
~.}
Comparing with (\ref{2.3}) we see that this solution corresponds to a Freund-Rubin ansatz, $F^{tot}_4\propto vol_4$, with $F^{tot}_0=0$ and $F^{tot}_2\propto J$.

\subsubsection*{Supersymmetry}

Let us now consider the supersymmetry of the solutions above. 
Imposing $H=0$ in addition to eqs.~(\ref{susysol}) implies: 
\beal\label{6.1}
e^{\phi} F_0=-\mathrm{Re}\left(\frac{c}{a}W\right)e^{-A}~;~~~~~ g_2=-\frac{72b}{a}We^{-A}
~.
\end{align}
Since $\phi$, $A$, $F_0$ are constant, it follows from the above that $c/a$ is constant. If $b\neq0$, 
the first equation in (\ref{constraint}) then implies that $W=0$ and consequently $AdS_4$ 
decompactifies to flat Minkowski space. If on the other hand $b=0$, the situation 
reduces to the rigid $SU(3)$ case, as follows from eq.~(3.5). The solution 
then falls within the class of supersymmetric $AdS_4$ solutions of \cite{lt}, from which it follows that supersymmetry 
enforces $F_0=0$. 

{\it In summary}: For nonzero Romans mass, the solutions presented in this section are not supersymmetric, as they 
violate the necessary conditions of section \ref{sec:solution}. For vanishing Romans mass  
there is an enhancement of supersymmetry, and these solutions 
fall within the class of the supersymmetric solutions of \cite{lt}.

\section{Static SU(2) structure in IIB}\label{sec:staticsu2}

It has been known for some time that static $SU(2)$-structure compactifications to $AdS_4$ are not allowed in IIA \cite{blt}. There is a  IIB counterpart of this no-go, forbidding strict $SU(3)$-structure compactifications to $AdS_4$ in IIB \cite{bciib}. However, static $SU(2)$-structure compactifications to $AdS_4$ {\it are} allowed in IIB. In this case 
we have $a=\pm b$, $c=0$, {\it cf.} eq.~(\ref{3.5}),  and  eqs.~(\ref{susysolb},\ref{constraintb}) simplify considerably to:
\boxedeq{\label{solstatic}
\spl{
f_1=f_2&=\pm 12iK^*\cdot\partial A \\
f_3&=6i (g_1+i g_2)\pm 12iW^*e^{-A}\\
h_1=h_2&=\pm 6i(g_1-i g_2)-12iW^*e^{-A}\\
h_3&=6i K^*\cdot\partial\left( 2A-\phi \right)
}
}
and
\boxedeq{\label{conststatic}
 g_2=\pm\frac{3i}{2} W^*e^{-A}~;~~~~~
a=\pm b=\mathrm{constant}\times e^{~\frac{1}{2}A}
~,
}
respectively. The $SU(2)$ structure, which can be read off off (\ref{23.1},\ref{21b},\ref{22b}), can be put in the  form:
\boxedeq{\label{23.1iib}
\spl{
dK_0&=-2W\mathrm{Im}~\!\omega_0\\
d\tilde{J}_0&=\mp g_0K_0\wedge\mathrm{Re}~\!\omega_0+\mathrm{c.c.}\\
d\omega_0&=\pm g_0K_0\wedge\widetilde{J}_0+\mathrm{c.c.}
~,
}}
where we have set: 
%
%
%
\eq{\label{defiib}
 K_0:=e^{3A-\phi}K~;~~~~~
\tilde{J}_0:= e^{2A-\phi}\tilde{J}~;~~~~~
\omega_0:= e^{2A-\phi}\omega~;~~~~~
g_0:=\frac{1}{2}e^{\phi-3A}\left(g_1\mp\frac{5}{2}W^*e^{-A}\right)
~.
}
Consistency requires that $d^2$  should annihilate $K_0$, $\widetilde{J}_0$, $\omega_0$, which is guaranteed 
provided $g_0$ is `holomorphic' ({\it cf.} the discussion around eq.~(\ref{4.1.1})):
\eq{\label{4.5}
dg_0=\frac{1}{2}K(K^*\cdot\partial g_0)
~.}
A special solution of the above is $g_0=\mathrm{constant}$.

\subsection*{Constant warp factor, dilaton}

A further simplification to eqs.~(\ref{solstatic}-\ref{23.1iib}) would be to assume constant 
dilaton and warp factor. Setting $\phi=A=0$ and demanding that 
$g_1$ be holomorphic, i.e. that it should satisfy the analogue of (\ref{4.5}), 
it is now straightforward to examine the Bianchi identities and equations of motion 
for all the form fields. Imposing $dH=0$ ({\it i.e.} demanding the absence of NS5 brane sources) implies:
\eq{\label{bih}
\mathrm{Re}(g_1W)=\frac{1}{3}\left(|g_1|^2+\frac{5}{4}|W|^2\right)
~,}
as follows from (\ref{b18},\ref{23.1iib}). Moreover we find a source (D7 branes/O7 planes) for the 
Bianchi identity of $F_1$:
\eq{\label{bi7}
dF_1=-\frac{4}{3}\left(|g_1|^2+\frac{5}{4}|W|^2\right)\mathrm{Im}~\!\omega~.
}
The source above corresponds to net orientifold charge. Note that demanding the absence of 
D7/O7 sets the cosmological constant to zero. 

In addition there is a potential source (D5 branes/O5 planes), which vanishes for special values of $g_1$, 
for the Bianchi identity of 
$F_3$:
\beal\label{bi5}
dF_3+H\wedge F_1=i\left(|g_1|^2-\frac{5}{4}|W|^2\right)\mathrm{Re}~\!\omega\wedge K\wedge K^*~.
\end{align}
There is a net orientifold charge for $|g_1|\geq\sqrt{5}/2|W|$.  
All other Bianchi's and equations of motion for the form-fields are automatically satisfied. It is then 
guaranteed by the integrability theorem of \cite{kt}, 
which generalizes the theorems of \cite{lt,gaun} 
to include calibrated sources, that all remaining equations of motion are automatically satisfied.

\subsection{Examples with smeared sources}\label{sec:smearediib}

In the following we will discuss two examples of supersymmetric IIB 
$AdS_4$ compactifications solving eqs.~(\ref{solstatic}-\ref{23.1iib}). 
Both of these examples, which have been mentioned before in the literature, contain sources smeared in the internal space.

\subsection*{Nilmanifold 5.1}

This example, where we take the internal six-dimentional manifold to be the nilmanifold 5.1, 
was first mentioned in \cite{klpt} and further examined in \cite{cetal}. The nilmanifold 5.1 
can be defined by specifying a coframe $e^i$, $i=1,\dots, 6$, such that:
\beal
de^i=0,~i=1,\dots, 5~;~~~~~de^6=e^{12}+e^{34} ~,
\end{align}
where $e^{ij}:=e^i\wedge e^j$. 
Let us set $A$, $\phi=0$ for simplicity. Moreover, assuming $a=+b$,  let us take 
\beal\label{ua}
g_1=\frac{5}{2}W^*~,
\end{align}
so that $g_0=0$, by virtue of (\ref{defiib}). Eqs.~(\ref{23.1iib}) are then satisfied, provided we identify:
\beal
K&=-2We^6+ie^5\nn\\
\widetilde{J}&=e^{13}-e^{24}\nn\\
\omega&=(ie^1+e^3)\wedge(ie^4+e^2)~.
\end{align}
This solution contains (smeared) O5/O7 sources, as can be seen by computing the 
 right-hand-sides of eqs.~(\ref{bi7},\ref{bi5}) above taking (\ref{ua}) into account.

\subsection*{T$^{1,1}\times$S$^1$}

In this example, which was first mentioned in \cite{klt}, we take the internal six-dimentional manifold to be the product  
$T^{1,1}\times S^1$. The total six-dimensional manifold admits a coset structure, decribed in section 4.6 of 
ref.~\cite{klt}, to which the reader is referred for further details\footnote{The present case corresponds to the 
$b=0$ embedding described in eqs.~(4.36,4.37) of ref.~\cite{klt}.}. As in the previous case, we can describe the internal manifold by specifying a coframe $e^i$, $i=1,\dots, 6$. The action of the exterior differential on the coframe 
is determined by the structure constants of the coset. As before, let us set $A$, $\phi=0$. Eqs.~(\ref{23.1iib}) are then satisfied, provided we identify:
\beal
K&=2We^3+ie^6\nn\\
\widetilde{J}&=-e^{14}+e^{25}\nn\\
\omega&=-i(ie^1+e^4)\wedge(ie^2-e^5)~.
\end{align}
In addition we must take $g_0=-1/2W\in\mathbb{R}$, so that:
\beal\label{ub}
g_1=-\frac{1}{W}+\frac{5}{2}W~.
\end{align}
As in the previous example, this solution contains (smeared) O5/O7 sources. It also generally 
contains (smeared) NS5-brane sources, 
which vanish for the special value: $W=\pm1/\sqrt{2}$, as can be seen from (\ref{bih},\ref{ub}).

\subsection{Examples with partially localized sources}\label{sec:iibloc}

Taking the limit to four-dimensional Minkowski space ($W\rightarrow0$), 
we will now discuss a class of supersymmetric IIB 
warped compactifications solving eqs.~(\ref{solstatic}-\ref{23.1iib}). 
These examples contain spacetime-filling NS5 and/or D5 branes partially localized in the internal space. 

Let us take $g_1=g_2=0$, so that $W=0$, in which case the external space becomes $\mathbb{R}^{1,3}$. 
It follows from (\ref{23.1iib}) that the two-forms $\big(e^{2A-\phi}\omega\big)$, 
$\big(e^{2A-\phi}\widetilde{J}\big)$ are closed, and therefore 
define a four-dimensional Calabi-Yau manifold, {\it i.e.} a $K3$ surface. 
It also follows from (\ref{23.1iib}) that the one-form $\big(e^{3A-\phi}K\big)$ is closed. We can 
therefore take it to be equal to $dz$, where  $z$ is a complex coordinate of a $T^2$.  

The metric of the six-dimensional internal manifold can therefore be written as:
\eq{\label{fmetric}
ds^2_6=e^{2\phi-6A}|d z|^2+ e^{\phi-2A}ds^2_{{K3}}~,
}
where $ds^2_{{K3}}$ is the metric of the $K3$ surface. Moreover, the non-zero fluxes can 
be read off off (\ref{b17},\ref{b18}):
\eq{\label{fflux}
\spl{
F_3&=-\frac{i}{2}\frac{\partial}{\partial z}\left(e^{-2A}\right)\mathrm{Re}~\!\omega_0\wedge dz+\mathrm{c.c.}\\
H&=-\frac{i}{2}\frac{\partial}{\partial z}\left(e^{\phi-2A}\right)\widetilde{J}_0\wedge dz+\mathrm{c.c.}
~.
}
}
As is now straightforward to compute, there will, in general, be source-terms in the Bianchi identities 
for the above form-fields, signalling the presence of NS5 and/or D5 branes. Indeed we find:
\eq{\spl{
dF_3&=\frac{i}{2}\frac{\partial^2}{\partial z\partial{z}^*}
\left(e^{-2A}\right)\mathrm{Re}~\!\omega_0\wedge dz\wedge dz^*+\mathrm{c.c.}\\
dH&=\frac{i}{2}\frac{\partial^2}{\partial z\partial{z}^*}
\left(e^{\phi-2A}\right)\widetilde{J}_0\wedge dz\wedge dz^*+\mathrm{c.c.}
~.
}}
Taking the functions $e^{-2A}$, $e^{\phi-2A}$ to be harmonic on $T^2$ ensures that the source-terms 
on the right-hand sides above are localized on $T^2$.

To complete the discussion of these solutions, 
one can also show that all remaining Bianchi identities and equations of motion for the form fields are 
satisfied for the system 
of fluxes given in (\ref{fflux}). As already remarked, the integrability theorem of \cite{kt} then guarantees that all remaining 
equations of motion will be automatically satisfied.

\section{Conclusions}

The scalar ansatz introduced in the present paper 
allowed us to explicitly 
solve the supersymmetry equations of type II supergravity. 
The `algebraic part' of the solution is given by eqs.~(\ref{susysol},\ref{constraint}) 
for IIA, and eqs.~(\ref{susysolb},\ref{constraintb}) for IIB. 
Moreover, these are necessary conditions which 
every supersymmetric $AdS_4$ solution should obey -- not only the solutions 
satisfying the scalar ansatz. In addition, eqs.~(\ref{4.0},\ref{3.8}) must be imposed in both cases. 

As already pointed out 
in the introduction, the algebraic part of the solution is unobstructed, as it simply 
expresses certain flux components in terms of a set of 
free parameters. The `differential part' of the solution is given in 
eq.~(\ref{23.1}), and specifies the local $SU(2)$ structure of the internal manifold.  
The main message of the present paper is therefore that: {\it in order to construct new supersymmetric 
$AdS_4$ compactifications of type II supergravity, it suffices to find six-dimensional 
manifolds which admit a local $SU(2)$ structure obeying eq.~(\ref{23.1}).} A natural direction 
for further study would be to systematically scan different classes of manifolds for that purpose.

Solutions of the supersymmetry equations will in general contain sources. The source content 
of a solution is revealed by studying the Bianchi identities of the form fields. As we have seen 
in the examples presented here, the sources present in a solution may or may not 
admit a satisfactory physical interpretation. At least one need not worry about the 
remaining equations of motion:  thanks 
to the integrability theorem of \cite{kt}, we know that these will be automatically satisfied.

The case of $AdS_4$ compactifications of IIB on manifolds of 
static $SU(2)$ structure is, in some sense, the analogue of the well-known strict-$SU(3)$ 
case in IIA. Nevertheless, it had not been systematically studied before. In section \ref{sec:staticsu2} 
 we examined this case in detail. In particular, eqs.~(\ref{solstatic},\ref{conststatic}) are necessary 
conditions that every supersymmetric $AdS_4$ solution of static $SU(2)$ structure should obey. 
The examples of solutions presented in section \ref{sec:smearediib}, had already 
appeared in the literature in \cite{klpt,klt}, whereas to our knowledge the example 
of section \ref{sec:iibloc} is new. The latter is obtained in the limit 
of four-dimensional Minkowski space, and contains partially localized NS5- and D5-branes. It is perhaps worth noting 
that this example does not fall into the GKP class \cite{gkp}.

The nonsupersymmetric solutions presented in 
section  \ref{eomsolutions} were anticipated by Romans already in \cite{roma}, 
although they only received a brief mention in that reference. As we have seen, 
these solutions 
naturally fall into three disctinct classes, eqs.~(\ref{2.10}-\ref{2.12}), the last 
of which can be thought of as a deformation of the Nilsson-Pope solution. 
The $CFT_3$ dual of the latter class was recently considered in \cite{gato}. It would 
be interesting to examine whether a $CFT_3$ dual can also be constructed for the other 
two classes.


\appendix


\section{Useful relations}\label{useful}

In this section we list the following relations which are useful in deriving the 
supersymmetry conditions of section \ref{sec:solution}. For a more complete list 
the reader may consult \cite{blt}.
\beal\label{a1}
H\eta_1^*&=-\frac{i}{3}h_2\eta_1-\frac{i}{6}h_3K_m\gamma^m\eta_1^*\nn\\
H\eta_2^*&=\frac{i}{3}h_3\eta_1-\frac{i}{6}h_1K_m\gamma^m\eta_1^*
\end{align}
\beal\label{a2}
H_m\eta_1&=\frac{i}{6}(h_3K_m+h_3^*K^*_m)\eta_1+\frac{1}{12}
(2h^*_2\widetilde{J}_{mn}-h_3^*\omega_{mn}-ih_1K_mK_n-ih_2^*K^*_mK_n)\gamma^n\eta_1^*\nn\\
H_m\eta_2&=\frac{i}{6}(h_2K_m+h_1^*K^*_m)\eta_1+\frac{1}{12}
(-2h^*_3\widetilde{J}_{mn}-h_1^*\omega_{mn}+ih_3K_mK_n+ih_3^*K^*_mK_n)\gamma^n\eta_1^*
\end{align}
and
\beal\label{a3}
e^{\phi}F\eta_1&=
\left\{f_0-\frac{1}{8}g_1-\frac{1}{12}g_3+i(f+f_1+\frac{1}{2}f_3)\right\}\eta_1
+(-\frac{i}{4}f_2+\frac{1}{24}g_2)K_m\gamma^m\eta_1^*
\nn\\
e^{\phi}F\eta_2&=(\frac{i}{2}f_2^*-\frac{1}{12}g_2^*)\eta_1
-\frac{1}{2}\left\{f_0-\frac{1}{8}g_1+\frac{1}{12}g_3+i(f+f_1-\frac{1}{2}f_3)\right\}K_m\gamma^m\eta_1^*
\end{align}
\beal\label{a4}
e^{\phi}F\gamma_m\eta^*_1&=(-\frac{i}{2}f_2^*+\frac{1}{12}g_2^*)K^*_m\eta_1\nn\\
&\!\!\!\! +\left\{
(f_1-f+\frac{i}{8}g_1+if_0)\widetilde{J}_{mn}+\frac{1}{2}(if+if_1-\frac{i}{2}f_3-\frac{1}{8}g_1+\frac{1}{12}g_3+f_0)K^*_mK_n \right\}\gamma^n\eta_1^*\nn\\
e^{\phi}F\gamma_m\eta^*_2&=(if+if_1+\frac{i}{2}f_3-\frac{1}{8}g_1-\frac{1}{12}g_3+f_0)K^*_m\eta_1\nn\\
&+\frac{1}{2}\left\{
(f_1-f+\frac{i}{8}g_1+if_0)\omega_{mn}+(-\frac{i}{2}f_2+\frac{1}{12}g_2)K^*_mK_n \right\}\gamma^n\eta_1^*
~,
\end{align}
for type IIA, while:
\beal\label{a5}
e^{\phi}F\eta_1&=
-\frac{i}{3}f_2^*\eta_1^*
+\left(g_1^*-ig_2^*+\frac{i}{6}f_3^*\right)K^*_m\gamma^m\eta_1
\nn\\
e^{\phi}F\eta_2&=
-2\left(g_1^*-ig_2^*-\frac{i}{6}f_3^*\right)\eta_1^*
+\frac{i}{6}f_1^*K^*_m\gamma^m\eta_1
\end{align}
\beal\label{a6}
e^{\phi}F\gamma_m\eta_1&=2\left(g_1+ig_2+\frac{i}{6}f_3\right)K_m\eta_1+\left\{
(g_2^*-ig_1^*)\omega_{mn}-\frac{i}{6}f_1K_mK_n \right\}\gamma^n\eta_1^*\nn\\
e^{\phi}F\gamma_m\eta_2&=\frac{i}{3}f_2 K_m\eta_1-\left\{
2(g_2^*-ig_1^*)\widetilde{J}_{mn}+\left(g_1+ig_2-\frac{i}{6}f_3\right)K_mK_n \right\}\gamma^n\eta_1^*
~,
\end{align}
for type IIB.

The relations above can be put in a slightly different form, which is sometimes more convenient, by 
making use of the identites:
\beal
\gamma_m\eta_1&=-\frac{i}{2}\omega_{mn}\gamma^n\eta_2+K_m\eta_2^*\nn\\
&=-i\widetilde{J}_{mn}\gamma^n\eta_1+K_m\eta_2^*
\end{align}
and
\beal
\gamma_m\eta_2&=-\frac{i}{2}\omega^*_{mn}\gamma^n\eta_1-K_m\eta_1^*\nn\\
&=i\widetilde{J}_{mn}\gamma^n\eta_2-K_m\eta_1^*
~,
\end{align}
which follow from the formul\ae{} of \cite{blt}. Taking the above into account we rewrite (\ref{a1},\ref{a2}) 
equivalently as:
\beal
H\eta_1&=\frac{i}{3}\left(-h_2^*\eta_1^*+h_3^*\eta_2^* \right)\nn\\
H\eta_2&=\frac{i}{3}\left( h_3^*\eta_1^*+h_1^*\eta_2^*\right)
\end{align}
and
\beal
H_m\eta_1&=\frac{i}{6}\left(h_3K_m\eta_1+h_1K_m\eta_2-h_2^*\gamma_m\eta_1^*+h_3^*\gamma_m\eta_2^* \right)\nn\\
H_m\eta_2&=\frac{i}{6}\left( h_2K_m\eta_1-h_3K_m\eta_2+h_3^*\gamma_m\eta_1^*+h_1^*\gamma_m\eta_2^*\right)
~,
\end{align}
and similarly for (\ref{a3})-(\ref{a6}).

\section{Tensor decompositions}\label{tensordecompositions}

For the tensor decompositions of the various 
fields with respect to the local 
 $SU(2)$ structure we follow closely \cite{blt}, to which the reader is referred for further details. 
In the case of the scalar ansatz the 
various formul\ae{} simplify considerably, and are listed in eqs.~(\ref{b16}-\ref{b18}).  

In terms of the local $SU(2)$ structure, the form fields decompose in general as follows. 

{\it Two-form}
\beal
e^{\phi}F_{mn}=f_{mn}+f_{[m}K_{n]}+f^*_{[m}K^*_{n]}+if_1K_{[m}K^*_{n]}~,
\end{align}
with
\beal
K^if_{im}=K^if_{i}=K^{*i}f_{i}=0~,
\end{align}
where $f_1$ is real. We further decompose
\beal
f_{mn}=\widetilde{f}_{mn}+\frac{1}{8}\omega_{mn}f_2+\frac{1}{8}\omega^*_{mn}f^*_2
+\frac{1}{4}\widetilde{J}_{mn}f_3~,
\end{align}
where $\widetilde{f}_{mn}$ is $(1,1)$ and traceless with respect to $\widetilde{J}_{mn}$, {\it i.e.} it 
transforms in the $\bf{3}$ of $SU(2)$.  
The scalar $f_2$ is complex whereas $f_3$ is real. 
Moreover, 
\beal
f_m=-\frac{1}{4}\omega_{m}{}^i\widetilde{f}_{1i}
-\frac{1}{4}\omega^*_{m}{}^i\widetilde{f}_{2i},
\end{align}
where $(\Pi^-)_m{}^n\widetilde{f}_{1n}= (\Pi^+)_m{}^n\widetilde{f}_{2n}=0$. {\it I.e.} 
$\widetilde{f}_{1i}$  transforms in the $\bf{2}$ of $SU(2)$ 
whereas $\widetilde{f}_{2i}$ transforms in the $\bf{\bar{2}}$.

{\it Three-form}
\beal
H_{mnp}=h_{mnp}+h_{[mn}K_{p]}+h^*_{[mn}K^*_{p]}+ih_{[m}K_{n}K^*_{p]}~,
\end{align}
with
\beal
K^ih_{imn}=K^ih_{im}=K^{*i}h_{im}=K^ih_{i}=0~,
\end{align}
where $h_m$ is real and $h_{mn}$ is complex. 
We further decompose
\beal
h_{mnp}=-\frac{3}{32}\omega^*_{[mn}\omega_{p]}{}^i\widetilde{h}_{1i}
-\frac{3}{32}\omega_{[mn}\omega^*_{p]}{}^i\widetilde{h}^*_{1i},
\end{align}
where $(\Pi^-)_m{}^n\widetilde{h}_{1n}=0$. 
Moreover
\beal
h_{mn}=\widetilde{h}_{mn}+\frac{1}{8}\omega_{mn}h_1+\frac{1}{8}\omega^*_{mn}h_2
+\frac{1}{4}\widetilde{J}_{mn}h_3~,
\end{align}
where $\widetilde{h}_{mn}$ is complex and  $(1,1)$ and traceless with respect to $\omega_{mn}$. 
The scalars $h_{1,2,3}$ are complex. 
Finally,
\beal
h_m=-\frac{1}{4}\omega_{m}{}^i\widetilde{h}_{2i}
-\frac{1}{4}\omega^*_{m}{}^i\widetilde{h}^*_{2i},
\end{align}
where $(\Pi^-)_m{}^n\widetilde{h}_{2n}=0$.

{\it Four-form}
\beal
e^{\phi}F_{mnpq}=g_{mnpq}+g_{[mnp}K_{q]}+g^*_{[mnp}K^*_{q]}+ig_{[mn}K_{p}K^*_{q]}~,
\end{align}
with
\beal
K^ig_{imnp}=K^ig_{imn}=K^{*i}g_{imn}=K^ig_{im}=0~,
\end{align}
where $g_{mnpq}$, $g_{mn}$ are real and $g_{mnp}$ is complex. 
We  further decompose
\beal
g_{mnpq}=\frac{3}{8}\omega_{[mn}\omega_{pq]}g_1,
\end{align}
where the scalar $g_1$ is real. 
Moreover
\beal
g_{mnp}=-\frac{3}{32}\omega^*_{[mn}\omega_{p]}{}^i\widetilde{g}_{1i}
-\frac{3}{32}\omega_{[mn}\omega^*_{p]}{}^i\widetilde{g}_{2i},
\end{align}
where $(\Pi^-)_m{}^n\widetilde{g}_{1n}=(\Pi^+)_m{}^n\widetilde{g}_{2n}=0$. 
Finally,
\beal
g_{mn}=\widetilde{g}_{mn}+\frac{1}{8}\omega_{mn}g_2+\frac{1}{8}\omega^*_{mn}g^*_2
+\frac{1}{4}\widetilde{J}_{mn}g_3~,
\end{align}
where $\widetilde{g}_{mn}$ is real and it is traceless with respect to $\omega_{mn}$. 
The scalar $g_{2}$ is complex whereas  $g_{3}$ is real.

{\it Six-form}
\beal
e^{\phi}F_{mnpqrs}=f\varepsilon_{mnpqrs}~.
\end{align}
For the tensor decompositions in IIB one proceeds in an analogous fashion.

\section{Local SU(2) structure}\label{sec:localstructure}

This appendix contains details of the derivation of eqs.~(\ref{23.1}). Moreover, 
at the end of the section 
we give the torsion classes of the $SU(3)$ structure specified by $(J^{(1)},\Omega^{(1)})$. 
A similar computation could be used to derive the torsion classes of the $SU(3)$ structure specified by 
$(J^{(2)},\Omega^{(2)})$.

Plugging the tensor decompositions (\ref{b16}-\ref{b18}) into 
the differential equations (\ref{susyeqsdiff}), taking the formul\ae{} in appendix 
\ref{useful} into account,  we obtain:
\beal\label{nablaeta1}
\nabla_m\eta_1=&-\partial_m\log{a}~\eta_1\nn\\
&+\left\{ K_m (K)_1+K^*_m(K^*)_1\right\}\eta_1 \nn\\
&+\left\{ \widetilde{J}_{mn}(\widetilde{J})_1
+\omega_{mn}(\omega)_1 
+K_{m}^*K_{n}(K^*K)_1
+K_{m}K_{n}(KK)_1\right\} \gamma^n\eta^*_1
~
\end{align}
and
\beal\label{nablaeta2}
\nabla_m\eta_2=&\frac{{c}}{{b}}\partial_m\log\frac{a}{c}~\eta_1-\partial_m\log {b}~\eta_2\nn\\
&+\left\{ K_m (K)_2+K^*_m(K^*)_2\right\}\eta_1 \nn\\
&+\left\{ \widetilde{J}_{mn}(\widetilde{J})_2
+\omega_{mn}(\omega)_2 
+K_{m}^*K_{n}(K^*K)_2
+K_{m}K_{n}(KK)_2\right\} \gamma^n\eta^*_1
~,
\end{align}
where
\beal\label{21}
(K)_1&:=  -\frac{i}{24}h_3 \nn\\
(K^*)_1&:= -\frac{i}{24}h_3^*+\frac{{b}}{8{a}}(if+if_1+\frac{i}{2}f_3-\frac{1}{8}g_1-\frac{1}{12}g_3+f_0) 
+\frac{{c^*}}{8{a}}(-\frac{i}{2}f_2^*+\frac{1}{12}g_2^*)
\nn\\
(\widetilde{J})_1&:= -\frac{1}{24}h_2^*+\frac{c^*}{8{a}}(f_1-f+if_0+\frac{i}{8}g_1) 
\nn\\
(\omega)_1&:=\frac{1}{48}h_3^*+\frac{{b}}{16{a}}(f_1-f+if_0+\frac{i}{8}g_1) 
\nn\\
(K^*K)_1&:= +\frac{i}{48}h_2^*+\frac{{b}}{8{a}}(-\frac{i}{4}f_2+\frac{1}{24}g_2)+\frac{{c^*}}{16{a}}(if+if_1-\frac{i}{2}f_3-\frac{1}{8}g_1+\frac{1}{12}g_3+f_0)
\nn\\
(KK)_1&:=\frac{i}{48}h_1
~
\end{align}
%
%
and
\beal\label{22}
(K)_2&:= \frac{i}{24}h_2+\frac{i{c}}{12{b}}h_3 \nn\\
(K^*)_2&:= \frac{i}{24}h_1^*+\frac{i{c}}{12{b}}h_3^*-\frac{{c}}{8{a}}(if+if_1+\frac{i}{2}f_3-\frac{1}{8}g_1-\frac{1}{12}g_3+f_0)  \nn\\
&~~~~~~~~~~~~~~~~~~~~~~+\frac{|{c}|^2}{8{a}{b}}(\frac{i}{2}f_2^*-\frac{1}{12}g_2^*)-\frac{{a}}{8{b}}(\frac{i}{2}f_2^*+\frac{1}{12}g_2^*)
\nn\\
(\widetilde{J})_2&:=\frac{c}{12b}h_2^*-\frac{1}{24}h_3^*-\frac{|{c}|^2}{8{a}{b}}(f_1-f+if_0+\frac{i}{8}g_1) -\frac{{a}}{8{b}}(f-f_1+i f_0+\frac{i}{8}g_1)
\nn\\
(\omega)_2&:=-\frac{1}{48}h_1^*-\frac{c}{24b}h_3^*-\frac{{c}}{16{a}}(f_1-f+if_0+\frac{i}{8}g_1)  
\nn\\
(K^*K)_2&:= \frac{i}{48}h_3^*-\frac{i{c}}{24{b}}h_2^*-\frac{{c}}{8{a}}(-\frac{i}{4}f_2+\frac{1}{24}g_2) -\frac{|{c}|^2}{16{a}{b}}(if+if_1-\frac{i}{2}f_3-\frac{1}{8}g_1+\frac{1}{12}g_3+f_0)\nn\\
&~~~~~~~~~~~~~~~~~~~~~~+\frac{{a}}{16{b}}(if+if_1-\frac{i}{2}f_3+\frac{1}{8}g_1-\frac{1}{12}g_3-f_0)\Big\} \nn\\
(KK)_2&:= -\frac{ic}{24b}h_1+\frac{i}{48}h_3
~
\end{align}
for type IIA. Similarly for IIB we have:
\beal\label{21b}
(K)_1&:=  -\frac{i}{24}h_3 -\frac{ib}{24a}f_2-\frac{c}{4a}(g_1+ig_2+\frac{i}{6}f_3)  \nn\\
(K^*)_1&:= -\frac{i}{24}h_3^*\nn\\
(\widetilde{J})_1&:= -\frac{1}{24}h_2^*+\frac{b}{4{a}}(g_2^*-ig^*_1) 
\nn\\
(\omega)_1&:=\frac{1}{48}h_3^*-\frac{c}{8{a}}(g_2^*-ig^*_1) 
\nn\\
(K^*K)_1&:= \frac{i}{48}h_2^*
\nn\\
(KK)_1&:=\frac{i}{48}h_1+\frac{b}{8a}(g_1+ig_2-\frac{i}{6}f_3)+\frac{ic}{48a}f_1
~
\end{align}
and
\beal\label{22b}
(K)_2&:= \frac{i}{24}h_2+\frac{i{c}}{12{b}}h_3 
+\frac{ic}{24a}f_2+\frac{c^2+a^2}{4ab}(g_1+ig_2)+\frac{i(c^2-a^2)}{24ab}f_3
\nn\\
(K^*)_2&:= \frac{i}{24}h_1^*+\frac{i{c}}{12{b}}h_3^* \nn\\
(\widetilde{J})_2&:=\frac{c}{12b}h_2^*-\frac{1}{24}h_3^*-\frac{c}{4a}(g_2^*-ig_1^*)
\nn\\
(\omega)_2&:=-\frac{1}{48}h_1^*-\frac{c}{24b}h_3^*+\frac{c^2+a^2}{8ab}(g_2^*-ig_1^*)
\nn\\
(K^*K)_2&:= \frac{i}{48}h_3^*-\frac{i{c}}{24{b}}h_2^*\nn\\
(KK)_2&:= -\frac{ic}{24b}h_1+\frac{i}{48}h_3-\frac{c}{8a}(g_1+ig_2-\frac{i}{6}f_3)-\frac{i(c^2-a^2)}{48ab}f_1
~.
\end{align}
It is now straightforward to read off the action of the exterior differential on the local structure. 
Plugging eqs.~(\ref{nablaeta1},\ref{nablaeta2}) into the definitions (\ref{3.54},\ref{3.55},\ref{3.6}), 
taking (\ref{2.87}) into account, we find:
\beal\label{23}
dK&=K^*\wedge K\left\{(K^*)_1-2(K^*K)_2 -\frac{1}{2}K\cdot\partial\log(ab)\right\}\nn\\
&+\omega\left\{-4(\omega)_2\right\}+\omega^*\left\{2(\widetilde{J})_1\right\}+\widetilde{J}\left\{-2(\widetilde{J})_2-4(\omega)_1\right\}
~.
\end{align}
\beal\label{24.1}
d\tilde{J}&=K\wedge\omega\left\{-2(K^*K)^*_1-i(\widetilde{J})^*_1 \right\}+K\wedge\omega^*\left\{-2(KK)_1-2i(\omega)^*_2 \right\}\nn\\
&+K\wedge\widetilde{J}\left\{(K)_1+(K^*)^*_1+2i(\omega)^*_1-i(\widetilde{J})^*_2 -K^*\cdot\partial\log a\right\}
+\mathrm{c.c}
~.
\end{align}
\beal\label{25.1}
d\omega&=
K\wedge\widetilde{J}\left\{2(KK)_1+(K^*)_2^*+4i(\omega)_2^*-\frac{c^*}{2b}K^*\cdot\partial\log\frac{c^*}{a}\right\}\nn\\
&+K^*\wedge\widetilde{J}\left\{2(K^*K)_1+(K)_2^*-2i(\widetilde{J})_1-\frac{c^*}{2b}K\cdot\partial\log\frac{c^*}{a}\right\}\nn\\
&+K\wedge\omega\left\{(K)_1-2(K^*K)^*_2-2i(\widetilde{J})^*_2-\frac{1}{2}K^*\cdot\partial\log(ab)\right\}\nn\\
&+K^*\wedge\omega\left\{(K^*)_1-2(KK)^*_2-4i(\omega)_1   -\frac{1}{2}K\cdot\partial\log(ab)\right\}
~.
\end{align}
The content of the three equations above is exactly equivalent to the content of the spinorial equations 
(\ref{nablaeta1},\ref{nablaeta2}).  Moreover we have:
\beal\label{26}
dJ^{(1)}&=K\wedge\omega\left\{-2(K^*K)^*_1-2i(\widetilde{J})^*_1 \right\}+K\wedge\omega^*\left\{-2(KK)_1 \right\}\nn\\
&+K\wedge\widetilde{J}\left\{(K)_1+(K^*)^*_1+4i(\omega)^*_1 -K^*\cdot\partial\log a\right\}
+\mathrm{c.c}
~.
\end{align}
\beal\label{27}
d\Omega^{(1)}&=K^*\wedge K\wedge \widetilde{J}\left\{-4i(K^*K)_1-2(\widetilde{J})_1  \right\}\nn\\
&+K^*\wedge\Omega^{(1)}\left\{2(K^*)_1-4i(\omega)_1-K\cdot\partial\log{a} \right\}+\widetilde{J}\wedge\widetilde{J}\left\{-4i(\widetilde{J})_1 \right\}
~.
\end{align}
\beal
dJ^{(2)}&=K\wedge\omega\left\{(K)_2-\frac{c}{2b}K^*\cdot\partial\log\frac{c}{a} \right\}
+K\wedge\omega^*\left\{(K^*)_2^*+4i(\omega)^*_2 -\frac{c^*}{2b}K^*\cdot\partial\log\frac{c^*}{a}\right\}\nn\\
&+K\wedge\widetilde{J}\left\{2(KK)_2+2(K^*K)^*_2+2i(\widetilde{J})_2^* +K^*\cdot\partial\log b\right\}
+\mathrm{c.c}
~.
\end{align}
\beal
d\Omega^{(2)}&=K^*\wedge K\wedge \widetilde{J}\left\{2i(K^*)_2+4(\omega)_2
-\frac{ic}{b}K\cdot\partial\log\frac{c}{a} \right\}\nn\\
&+K^*\wedge K\wedge\omega^*\left\{-4i(K^*K)_2-2(\widetilde{J})_2-iK\cdot\partial\log{b} \right\}
+\widetilde{J}\wedge\widetilde{J}\left\{-8i(\omega)_2 \right\}
~.
\end{align}
It is also useful to define:
\begin{align}\label{3.57}
\widetilde{\Omega}_{mnp}:=\eta_2\gamma_{mnp}\eta_1~,
\end{align}
so that:
\beal\label{3.88}
\widetilde{\Omega}=i\widetilde{J}\wedge K~.
\end{align}
We find:
\beal
d\tilde{\Omega}&=K^*\wedge K\wedge \widetilde{J}\left\{i(K^*)_1-2i(K^*K)_2+2(\omega)_1
-(\widetilde{J})_2-\frac{i}{2}K\cdot\partial\log(ab)  \right\}\nn\\
&+K^*\wedge\Omega^{(1)}\left\{(K^*)_2-2i(\omega)_2-\frac{c}{2b}K\cdot\partial\log\frac{c}{a} \right\}\nn\\
&+\widetilde{J}\wedge\widetilde{J}\left\{-4i(\omega)_1-2i(\widetilde{J})_2 \right\}
+K^*\wedge K\wedge\omega^*\left\{-2i(K^*K)_1-(\widetilde{J})_1 \right\}
~.
\end{align}
One can perform several consistency checks of these expressions. For example, $d\Omega^{(1)}$ can be computed in 
two different ways: either directly  by plugging eq.~(\ref{nablaeta1}) into definition (\ref{3.54}), or by plugging the 
expressions for $d\omega$, $dK$ above into $d\Omega^{(1)}=-i d\omega\wedge K-i\omega\wedge dK$, which follows from eq.~(\ref{2.87}). In order to perform these consistency checks, it is useful to take the following equations into account:
\beal\label{pirto}
(K)_1+(K^*)^*_1&=K^*\cdot\partial\log{a}\nn\\
(K^*K)_2+(KK)_2^*&=-\frac{1}{2}K\cdot\partial\log{b}\nn\\
(K^*)_2-2(KK)_1^*&=\frac{c}{2b}K\cdot\partial\log\frac{c}{a}\nn\\
(K)_2-2(K^*K)_1^*&=\frac{c}{2b}K^*\cdot\partial\log\frac{c}{a}\nn\\
d\log\frac{|c|}{a}&=-\frac{b^2}{|c|^2}d\log\frac{b}{a}
~.
\end{align}
The first four equations above can be shown to be equivalent to $\mathcal{L}_K(\eta_i^{\dagger}\eta_j)=\mathcal{L}_{K^*}(\eta_i^{\dagger}\eta_j)=0$, for $i,j=1,2$, once (\ref{nablaeta1},\ref{nablaeta2}) are taken into account.  The 
last relation follows from (\ref{3.8}). 
Alternatively, eqs.~(\ref{pirto}) 
can be derived directly from the solution (\ref{susysol},\ref{3.8}) and the constraints (\ref{constraint}) in IIA, 
and similarly in IIB.

\subsubsection*{Torsion classes}

As discussed in some detail in section 
\ref{sec:supersymmetrya}, 
each of the two spinors $\theta_{1,2}$ can be used to define an $SU(3)$ structure on $\mathcal{M}_6$. 
On the other hand, for an $SU(3)$-structure manifold, the torsion classes are defined via:
\beal
dJ&=\frac{3i}{4}\left(\mathcal{W}_1\Omega^*-\mathcal{W}^*_1\Omega\right)+\mathcal{W}_3+\mathcal{W}_4\wedge J\nn\\
d\Omega&=\mathcal{W}_1 J\wedge J+\mathcal{W}_2\wedge J+\mathcal{W}_5^*\wedge\Omega~.
\end{align}
In particular, the torsion classes corresponding to the $SU(3)$ structure  $(J^{(1)},\Omega^{(1)})$ can be read off by 
comparing the above with (\ref{26},\ref{27}), taking (\ref{pirto}) into account:
\eq{\label{torcla}
\spl{
\mathcal{W}_1&=-\frac{8i}{3}\left\{(\widetilde{J})_1+i(K^*K)_1\right\}\\
\mathcal{W}_2&=-\frac{4i}{3}\big(J^{(1)}-\frac{3i}{2} K\wedge K^*\big)\left\{(\widetilde{J})_1-2i(K^*K)_1\right\}\\
\mathcal{W}_3&=K\wedge\omega^*\left\{-2(KK)_1\right\}+\mathrm{c.c.}\\
\mathcal{W}_4&=K\left\{4i(\omega)_1^*\right\}+\mathrm{c.c.}\\
\mathcal{W}_5&=K\left\{(K^*)^*_1-(K)_1+4i(\omega)_1^*\right\}
~.}}
%
We see that $\mathcal{W}_{5}$ is proportional to $K$. 
Moreover, in the IIA case,  taking (\ref{susysol},\ref{21}) into account 
we find that $\mathcal{W}_{4}$ is exact: $\mathcal{W}_{4}=d(\phi-3A)$. Therefore $\mathcal{W}_{4}$ can 
be removed by a conformal rescaling of the internal metric: $ds^2_6\rightarrow e^{3A-\phi}ds^2_6$.

%
%

\section{Sasaki-Einstein}\label{sec:sasaki}

There is a well-known class of eleven-dimensional supergravity solutions of the form 
$\ads\times \mcal_7$, where $\mcal_7$ is a seven-dimensional Einstein manifold. 
Specifically, the eleven - dimensional 
metric is given by
\al{\label{1.1}
ds^2=ds^2(\ads)+ds^2(\mcal_7)~,
}
while the four-form flux is of Freund-Rubin type: $G_4\propto vol_4$, where 
$vol_4$ is the volume form of $\ads$. In addition, the manifold $\mcal_7$ has the 
property that the cone over it, $\ccal(\mcal_7)$,  is an eight-dimensional manifold of special holonomy. 
The supersymmetry preserved by the solution depends on the holonomy of $\ccal(\mcal_7)$. Table 
\ref{holo} lists the type of 
\begin{table}[h!]
\begin{center}
\begin{tabular}{|c|c|c|}
\hline
$\mcal$ & $\mathcal{H}ol(\ccal(\mcal_7))$ & $\ncal$\\
\hline
Weak $G_2$ & $Spin(7)$ & 1 \\
Sasaki-Einstein & $SU(4)$ & 2\\
tri-Sasaki & $Sp(2)$ & 3\\
$S^7$ & 1 & 8\\
\hline
\end{tabular}
\caption{List of seven-dimensional Einstein manifolds $\mcal_7$, the holonomy of the 
corresponding eight-dimensional cones and 
the number of preserved supersymmetries in four dimensions.}
\label{holo}
\end{center}
\end{table}
the seven-dimensional Einstein manifold $\mcal_7$, the holonomy of the cone over it, $\mathcal{H}ol(\ccal(\mcal_7))$, as well as the 
number of preserved supersymmetries, $\ncal$, in four dimensions.

We will now specialize to the case where $\mathcal{H}ol(\ccal(\mcal_7))$ is a subgroup of $SU(4)$, {\it i.e.} 
the eight-dimensional cone is Calabi-Yau. Equivalently, we will take $\mcal_7$ to be Sasaki-Einstein (which includes 
the $S^7$ and the tri-Sasaki as special cases). The manifold $\mcal_7$ can then be thought of as the total space 
of a fibre bundle with connection one-form $\mathcal{A}$ on a six-dimensional base-space $\mcal_6$,
\al{\label{1.2}
ds^2(\mcal_7)=(dy+\mathcal{A})^2+ds^2(\mcal_6)
~,}
where $ds^2(\mcal_6)$ is a local K\"{a}hler-Einstein metric and $y$ is the coordinate on the fibre. The Killing vector 
$\partial_y$ is the so-called `Reeb vector'. If the  orbits of the Reeb vector are closed and the $U(1)$ action is free, $\mcal_7$ is regular and 
$\mcal_6$ is globally a manifold. 
One can define a local $SU(3)$ structure on $\mcal_6$ specified by a K\"{a}hler form $J$ and a complex three-form $\Omega$, 
such that $d\mathcal{A}=2J$ and $d\Omega=4i\mathcal{A}\wedge\Omega$. Note, however, that globally the structure group of $\mcal_6$ is  not $SU(3)$ but rather $U(3)$, since $\Omega$ is not globally defined in general.

A useful property of odd-dimensional, simply-connected 
Sasaki-Einstein manifolds is that they admit at least two Killing spinors. In the seven-dimensional case,
 it was shown in \cite{frie} that, under certain regularity 
assumptions, the converse is also true: any pair of (real) Killing spinors defines 
a Sasaki-Einstein structure on $\mcal_7$. Moreover, there is a one-to-one  correspondence between 
triplets of Killing spinors and tri-Sasaki structures on $\mcal_7$.


\begin{thebibliography}{99}



\bibitem{blg}
 J.~Bagger and N.~Lambert,
 ``Modeling multiple M2's,''
 \prd{75}{2007}{045020}
 [arXiv:\hepth{0611108}];
 A.~Gustavsson,
 ``Algebraic structures on parallel M2-branes,''
 \arXividhepth{0709.1260};
 J.~Bagger and N.~Lambert,
 ``Gauge symmetry and supersymmetry of multiple M2-Branes,''
 \prd{77}{2008}{065008}
 [\arXividhepth{0711.0955}];
  J.~Bagger and N.~Lambert,
 ``Comments On Multiple M2-branes,''
 \jhep{0802}{2008}{105}
 [\arXividhepth{0712.3738}].



\bibitem{abjm}
  O.~Aharony, O.~Bergman, D.~L.~Jafferis and J.~Maldacena,
  ``N=6 superconformal Chern-Simons-matter theories, M2-branes and their
  gravity duals,''
  \jhep{0810}{2008}{091} 
  [arXiv:0806.1218 [hep-th]].










\bibitem{lt}
 D.~L\"ust and D.~Tsimpis,
 ``Supersymmetric AdS$_4$ compactifications of IIA supergravity,''
 \jhep{0502}{2005}{027}
 [arXiv:\hepth{0412250}].




\bibitem{np}
 B.~E.~W.~Nilsson and C.~N.~Pope,
 ``Hopf Fibration Of Eleven-Dimensional Supergravity,''
 \cqg{1}{1984}{499}.

\bibitem{npa}
  D.~P.~Sorokin, V.~I.~Tkach and D.~V.~Volkov,
  ``Kaluza-Klein Theories And Spontaneous Compactification Mechanisms Of Extra
  Space Dimensions,''
{\it  In *Moscow 1984, Proceedings, Quantum Gravity*, 376-392}

\bibitem{npb}
  D.~P.~Sorokin, V.~I.~Tkach and D.~V.~Volkov,
  ``On The Relationship Between Compactified Vacua Of D = 11 And D = 10
  Supergravities,''
  \plb{ 161}{1985}{301}.


\bibitem{font}
 G.~Aldazabal and A.~Font,
 ``A second look at $\mathcal{N}=1$ supersymmetric AdS$_4$ vacua of type IIA supergravity,''
 \jhep{0802}{2008}{086} [\arXividhepth{0712.1021}].








\bibitem{bc}
 K.~Behrndt and M.~Cveti\v{c},
 ``General $\mathcal{N} = 1$ supersymmetric flux vacua of (massive) type IIA string theory'',
 \prl{95}{2005}{021601} [arXiv:\hepth{0403049}];
 ``General $\mathcal{N} = 1$ supersymmetric fluxes in massive type IIA string theory'',
 \npb{708}{2005}{45} [arXiv:\hepth{0407263}].












\bibitem{toma}
 A.~Tomasiello,
 ``New string vacua from twistor spaces,''
 \arXividhepth{0712.1396}.

\bibitem{klt}
 P.~Koerber, D.~L\"{u}st and D.~Tsimpis,
 ``Type IIA AdS$_4$ compactifications on cosets, interpolations and domain
 walls,''
 \jhep{0807}{2008}{017} [\arXividhepth{0804.0614}].





\bibitem{bciib}
  K.~Behrndt, M.~Cvetic and P.~Gao,
  ``General type IIB fluxes with SU(3) structures,''
  \npb{721}{2005}{287} 
  [arXiv:hep-th/0502154].




































\bibitem{scan}
 M.~Gra\~{n}a, R.~Minasian, M.~Petrini and A.~Tomasiello,
 ``A scan for new ${\cal N}=1$ vacua on twisted tori,''
 \jhep{0705}{2007}{031}
 [arXiv:\hepth{0609124}].
 


\bibitem{tsim}
  D.~Tsimpis,
  ``M-theory on eight-manifolds revisited: N = 1 supersymmetry and  generalized
  Spin(7) structures,''
  \jhep{0604}{2006}{027}
  [\arXividhepth{0511047}].


\bibitem{andriot}
 D.~Andriot,
 ``New supersymmetric flux vacua with intermediate SU(2) structure,''
 \arXividhepth{0804.1769}.


\bibitem{cetal}
 C.~Caviezel, P.~Koerber, S.~Kors, D.~L\"{u}st, D.~Tsimpis and M.~Zagermann,
 ``The effective theory of type IIA AdS4 compactifications on nilmanifolds and
 cosets,''
 arXiv:0806.3458 [hep-th].


\bibitem{blt}
 J.~Bovy, D.~L\"{u}st and D.~Tsimpis,
 ``N = 1,2 supersymmetric vacua of IIA supergravity and SU(2) structures,''
 \jhep{0508}{2005}{056} 
 [arXiv:\hepth{0506160}].



\bibitem{gene}
 M.~Gra\~{n}a, R.~Minasian, M.~Petrini and A.~Tomasiello,
 {\em Generalized structures of ${\cal N} = 1$ vacua},
 \jhep{0511}{2005}{020} [arXiv:\hepth{0505212}].




\bibitem{kt}
 P.~Koerber and D.~Tsimpis,
 ``Supersymmetric sources, integrability and generalized-structure
 compactifications,''
 \jhep{0708}{2007}{082}
 [ \arXividhepth{0706.1244}].





\bibitem{roma}
  L.~J.~Romans,
  ``Massive N=2a Supergravity In Ten-Dimensions,''
  \plb{169}{1986}{374}.



\bibitem{gato}
  D.~Gaiotto and A.~Tomasiello,
  ``The gauge dual of Romans mass,''
  arXiv:0901.0969 [hep-th].



\bibitem{lmmt}
 D.~L\"{u}st, F.~Marchesano, L.~Martucci and D.~Tsimpis,
 ``Generalized non-supersymmetric flux vacua,''
  \arXividhepth{0807.4540}.




\bibitem{zafa}
  R.~Minasian, M.~Petrini and A.~Zaffaroni,
  ``Gravity duals to deformed SYM theories and generalized complex geometry,''
  \jhep{0612}{055}{2006}
  [\arXividhepth{0606257}].




\bibitem{hato}
  N.~Halmagyi and A.~Tomasiello,
  ``Generalized Kaehler Potentials from Supergravity,''
 \arXividhepth{0708.1032}.


\bibitem{andrun}
D.~Andriot, unpublished.

\bibitem{neu}
  I.~P.~Neupane,
  ``Simple cosmological de Sitter solutions on dS$_4 \times Y_6$ spaces,''
  arXiv:0901.2568 [hep-th].


\bibitem{gaun}
  J.~P.~Gauntlett, D.~Martelli, J.~Sparks and D.~Waldram,
  ``Supersymmetric AdS(5) solutions of type IIB supergravity,''
  \cqg{23}{2006}{4693} 
  [arXiv:hep-th/0510125].




\bibitem{klpt}
  C.~Kounnas, D.~L\"{u}st, P.~M.~Petropoulos and D.~Tsimpis,
  ``AdS4 flux vacua in type II superstrings and their domain-wall solutions,''
  \jhep{0709}{2007}{051} 
  [arXiv:0707.4270 [hep-th]].



\bibitem{frie}
 T.~Friedrich and I.~Kath,
 ``Seven-dimensional compact Riemannian manifolds with Killing spinors,''
 \cmp{133}{1990}{543}.



\bibitem{gkp}
  S.~B.~Giddings, S.~Kachru and J.~Polchinski,
  ``Hierarchies from fluxes in string compactifications,''
  \prd{\bf 66}{2002}{106006}
  [arXiv:hep-th/0105097].

\end{thebibliography}
\end{document}